\documentstyle[preprint,aps,floats,psfig]{revtex}
\begin{document}
%
%my defs:
\def\mh{m_h^{}}
\def\vev#1{{\langle#1\rangle}}
\def\gev{{\rm GeV}}
\def\tev{{\rm TeV}}
\def\fbi{\rm fb^{-1}}
%\def\lsim{\mathrel{\raise.3ex\hbox{$<$\kern-.75em\lower1ex\hbox{$\sim$}}}}
%\def\gsim{\mathrel{\raise.3ex\hbox{$>$\kern-.75em\lower1ex\hbox{$\sim$}}}}
%%%%%%%%%%%%%%%%%%%%%%%%%%%%%%%%
%%%%%%%%  Slash character...

\newcommand{ \slashchar }[1]{\setbox0=\hbox{$#1$}   % set a box for #1
   \dimen0=\wd0                                     % and get its size
   \setbox1=\hbox{/} \dimen1=\wd1                   % get size of /
   \ifdim\dimen0>\dimen1                            % #1 is bigger
      \rlap{\hbox to \dimen0{\hfil/\hfil}}          % so center / in box
      #1                                            % and print #1
   \else                                            % / is bigger
      \rlap{\hbox to \dimen1{\hfil$#1$\hfil}}       % so center #1
      /                                             % and print /
   \fi}                                             %

\tighten
\preprint{ \vbox{
\hbox{MADPH--99-1123}
\hbox{hep-ph/9906508}}}
\draft
\title{A gauge-mediated supersymmetry breaking model\\ 
with an extra singlet Higgs field}
\author{Tao Han, Danny Marfatia and Ren-Jie Zhang}
\address{Department of Physics, University of Wisconsin\\ 
1150 University Avenue, Madison, WI 53706, USA}
\date{August 1999}

\maketitle

\begin{abstract}
We study in some detail the next-to-minimal supersymmetric
 standard model with gauge mediation of supersymmetry breaking. 
We find that it is feasible to spontaneously generate values 
of the Higgs mass parameters
$\mu$ and $B_\mu$ consistent with radiative electroweak 
symmetry breaking. The model has a phenomenologically viable particle
spectrum. Messenger sneutrinos with mass
in the range \mbox{6 to 25 TeV} can serve as cold dark matter.
It is also possible to evade the cosmological domain wall problem 
in this scenario.
\end{abstract}
\pacs{}
%\pacs{14.80.Bn, 13.85.Qk}

\section{Introduction}

The Standard Model (SM) fulfills electroweak symmetry breaking 
(EWSB) by introducing a scalar Higgs particle. 
The mass-squared of this particle suffers 
from quadratically divergent radiative corrections, and is therefore
sensitive to physics effects in the ultraviolet regime, possibly
at the grand unification scale ($M_{GUT}$) or the Planck scale ($M_{Pl}$).
Stabilizing the electroweak scale against radiative corrections
is the primary motivation for exploring theories beyond the SM, 
among which weak-scale supersymmetry (SUSY) has been the
leading candidate \cite{rep}.

In the minimal supersymmetric extension of the standard model (MSSM), 
the Higgs scalar potential has the form
\begin{equation}
V \supset (\mu^2+m_{H_u}^2)\ |H_u|^2 +(\mu^2+m_{H_d}^2)\ |H_d|^2 
- ( B_{\mu}{H_u}{H_d}+h.c.)+\ldots
\end{equation}
where ${H_u}$ and ${H_d}$ are two Higgs doublets coupled to the up and
down type fermions respectively, 
$m_{H_{u,d}}$ and $B_\mu$ are soft SUSY
breaking parameters, and $\mu$ is a mass parameter in the superpotential
\begin{equation}
W_\mu \supset \mu{H_u}{H_d}.
\end{equation}
To achieve correct electroweak symmetry breaking without
drastic fine-tuning, all these mass 
parameters should have the size of ${\cal O}(M_Z)$.
The soft parameters can be made to satisfy this requirement. 
The $\mu$ parameter on the other hand, should naively be of the 
order of the fundamental scale of the theory like $M_{Pl}$ or
$M_{GUT}$ since it is a dimensionful parameter of the superpotential. 
This apparent mismatch is termed as the SUSY $\mu$-problem.

There are many approaches to resolving the $\mu$-problem \cite{NMSSM,GM}, 
the simplest one being the introduction of an extra singlet 
Higgs field $N$. The Higgs superpotential is then written as
\begin{equation}
   W_N \supset {\lambda_N}N{H_u}{H_d}-{\frac{k}{3}}{N^3}.
\label{nmssm}
\end{equation}
This is called the next-to-minimal supersymmetric 
standard model (NMSSM) \cite{NMSSM}. The bilinear term 
$\mu{H_u}{H_d}$ is absent by invoking a discrete $\mathbf{Z_3}$ 
symmetry under which every chiral superfield 
$\Phi$ transforms as $\Phi \rightarrow e^{2\pi i/3}\Phi$.
When $N$ acquires a vacuum expectation value (VEV) 
$\vev{N}\approx {\cal O}(M_Z)$, 
the $\mu$-parameter is effectively generated as
\begin{equation}
\mu=\lambda_{N}\vev{N}\sim {\cal O}(M_Z).
\end{equation}
The inclusion of the $N^{3}$ term in Eq.~(\ref{nmssm}) is 
important since without it, the superpotential would have a 
problematic Peccei-Quinn symmetry. The spontaneous breaking 
of the $\mathbf{Z_3}$ symmetry by $\vev{N}$ leads to the well-known 
cosmological domain wall problem \cite{Domain}. Also,
one should be aware of the potential problem of destabilizing
the electroweak scale from tadpoles associated with a gauge 
singlet \cite{singlet}.

In gauge-mediated supersymmetry breaking models \cite{xis,Dine,GMSB},
the SUSY breaking in the secluded sector is usually parameterized by  
a SM singlet chiral superfield $S$ with non-vanishing 
VEVs for the lowest and highest components 
$\langle S\rangle+\theta^2 F_S$.
If the superpotential contains the term
\begin{equation}
   W_S \supset {\lambda_S}S{H_u}{H_d}.
\label{ws}
\end{equation}
then $\mu=\lambda_S\langle S\rangle$ and $B_\mu=
\lambda_S F_S$, which implies
\begin{equation}
B_\mu\ \simeq\ \mu\Lambda\ ,
\label{mubmu}
\end{equation}
where $\Lambda=F_S/\langle S\rangle\sim {\cal O} (100$ TeV).
Therefore $\mu$ and $\sqrt{B_\mu}$ 
cannot both be ${\cal O} (M_Z)$.
This is known as the $\mu$-problem in GMSB.
Some solutions have been proposed in \cite{mugmsb,Murayama}. 
In particular, Ref. \cite{Murayama} studied the GMSB paradigm
with the NMSSM,  but the conclusion is negative if one does not introduce
extra vector-like quarks in the theory.

In this paper we perform a detailed analysis of a gauge-mediated
model with an extra singlet Higgs field in a most general
superpotential including all couplings between the singlets, 
messengers and Higgs fields, respecting the $\mathbf{Z_3}$ symmetry. 
Without constructing a SUSY breaking model explicitly, 
we consider it as a phenomenological model and study its 
consistency with theoretical and experimental
constraints. 
In Section II we describe our general framework for the 
gauge-mediated SUSY breaking model with an extra singlet.
In Section III we present a numerical study of this model,
concentrating on the $\mu$-problem and phenomenologically
viable mass spectrum of SUSY particles (sparticles) 
with emphasis on the next-to-the-lightest supersymmetric particle
(NLSP) and Higgs bosons. 
In contrast to the conclusion of \mbox{Ref.~\cite{Murayama},} 
we find that suitable $\mu$ and $B_\mu$ parameters can be generated 
in this minimal scenario with all phenomenological constraints satisfied.
In Section IV we show that messenger sneutrinos can serve as a
cold dark matter candidate and that they are sufficiently massive to 
evade the current experimental bound from direct searches.
In \mbox{Section V} we demonstrate that the domain wall problem 
may find a solution due to fast decay of the walls
induced by higher dimensional operators, yet consistent with EWSB.
 We conclude in 
Section VI. Some technical details of the model---the renormalization
group equations (RGEs) and Higgs boson
mass matrices---are provided in two appendices.

\section{The minimal GMSB with an extra singlet}

In our phenomenological approach to the GMSB, we parameterize 
SUSY breaking in the secluded sector by a SM gauge singlet spurion
$S=\langle S\rangle+\theta^2 F_S$.
The SUSY breaking is then transmitted to the observable
sector via vector-like messengers $\Phi$ and $\overline\Phi$ 
which couple to $S$ according to 
$S\overline\Phi\Phi$. In the simplest case,  $\Phi$ and $\overline\Phi$ 
transform as a single flavor of $\mathbf{5+\overline 5}$ of SU(5).

In gauge-mediated models not involving any direct messenger-matter 
interactions, it is not possible to generate  $\mu$ and $B_\mu$
 consistent with EWSB.
 In fact, even  if the spurion $S$ couples directly to 
the Higgs superfields as in Eq.~(\ref{ws}),
one can easily show that $\mu=\lambda_S\langle S\rangle$ and 
$B_\mu=\lambda_S F_S$, which leads to the troublesome relation
in Eq.~(\ref{mubmu}). This is a general result irrespective of whether $\mu$
is generated spontaneously or radiatively.  
 One cannot get both $\mu$ and $\sqrt{B_\mu}$ 
of the order of $M_Z$ and EWSB fails because the large value of $B_\mu$ 
destabilizes the scalar potential. 

Motivated by the NMSSM model, we consider a minimal gauge-mediated 
model with an extra singlet Higgs field. The most general 
superpotential respecting a $\mathbf{Z_3}$ symmetry is
\begin{equation}
W=\xi_{S}S\overline\Phi\Phi+\xi_{N}N\overline\Phi\Phi
-{\frac{\eta_{S}}{2}}{S^2}N
-{\frac{\eta_{N}}{2}}{N^2}S+{{\lambda }_S}S{H_u}{H_d}+
{{\lambda}_N}N{H_u}{H_d}-{\frac{k}{3}}{N^3}. 
\label{model}
\end{equation}
The gauginos and scalars acquire mass radiatively at the 1-loop and 2-loop 
level respectively,  and have the following form 
\begin{eqnarray}
M_{i} &\simeq& k_{i} {\frac{\alpha_{i}}{4\pi}}{\Lambda} \label{gaugino},\\
m^{2}_{\tilde{f}} &\simeq& 2\sum_{i=1}^{3} {C^{\tilde{f}}_{i} k_{i} 
({\frac{\alpha_{i}}{4\pi}}{\Lambda})^{2}}\ ,
\label{msq}
\end{eqnarray}
where \(k_{1}=5/3, k_{2}=k_{3}=1\), 
and $C^{\tilde{f}}_{1}=Y^{2}$ for U(1),  
$C^{\tilde{f}}_{2}=3/4$ for weak SU(2) 
doublets and $C^{\tilde{f}}_{3}=4/3$ for color triplets. 
$\Lambda=F_{S}/S$ is the effective SUSY-breaking scale. 
In models in which messengers and matter do not interact directly,
 the trilinear soft $A$-terms and $m^{2}_{N}$ arise 
only at 2-loop and 3-loop respectively, and can be neglected.
The superpotential Eq.~(\ref{model}) contains messenger-matter couplings,
 thus inducing $A_{\lambda,k}$ and $m^{2}_{N}$ at 1-loop and 2-loop
 respectively \cite{GMSB},
\begin{eqnarray}
{\frac{A_k}{3}} &=&A_{\lambda_N}\simeq  -5{\frac{\alpha_{\xi_{N}}}
{4\pi}}{\Lambda},\\
m^{2}_{N} &\simeq& 10 {\alpha_{\xi_N}} \left({\frac{7}{2}{\alpha_{\xi_N}}}
-{\frac{8}{5}}{\alpha_3}-{\frac{3}{5}}{\alpha_2}-{\frac{13}{3}}{\alpha_1}\right)
 \left({\frac{\Lambda}{4\pi}}\right)^2,
\end{eqnarray}
where $\alpha_{\xi_N}={\xi_N^2}/4\pi$. Note that the
contributions from the gauge couplings can make  $m^{2}_{N}$ negative for
small values of $\xi_N$.
Among the couplings in Eq.~(\ref{model}), we may anticipate that 
$\xi_{S},\xi_{N},\lambda_N,k\sim {\cal O}(1)$ and that  $\eta_N$,
 $\eta_S$ and $\lambda_S$ will be small due to the 
large VEV from $S$. We will determine the pattern
for phenomenologically viable solutions. The only result contrary to these
 expectations is that we find $\eta_N\sim {\cal O}(1)$.  
 We will not explore the possible symmetries 
in terms of model-building that naturally lead to this 
pattern \cite{xis,Dine,Murayama}. 

The relevant part of the scalar potential is
\begin{equation}
 V_{Higgs}= V_{F}+ V_{D}+ V_{tadpole}+ V_{soft}+\Delta V,
\label{pot}
\end{equation}
where
\begin{eqnarray}
&&V_{F}\ =\ \mid{\lambda }_SS+{\lambda }_N 
N\!\mid{^2}(\,\mid\!H_{u}\!\mid^{2}+\mid\!H_{d}\!\mid^{2})+ \mid\!{\lambda 
}_N{H_u}{H_d}-k{N^2-{\frac{\eta_{S}}{2}}{S^2}}-\eta_{N}NS\!\mid^{2}\nonumber\\
&&\qquad+\mid\!{\lambda}_S{H_u}{H_d}+F_{S}-{\frac{\eta_{N}}{2}}{N^2}
-\eta_{S}NS\!\mid^{2},\\
&&V_{D}\ =\ {\frac{g_{2}^{2}}{8}}(H_{u}^{\dagger}
\,\vec{\sigma}\,H_{u}+H_{d}^{\dagger}
\,\vec{\sigma}\,H_{d})^{2}+{\frac{g'^{\,2}}{8}}
(\,\mid\!H_{u}\!\mid^{2}-\mid\!H_{d}\!\mid^{2})^{2},\\
&&V_{tadpole}\ =\ {\frac{\xi_{S}\xi_{N}}{8\pi^{2}}}{\frac{NF_{S}^{2}}{S}},\\
&&V_{soft}\ =\ m^{2}_{H_{u}}\mid\!H_{u}\!\mid^{2}
+\,m^{2}_{H_{d}}\mid\!H_{d}\!\mid^{2}
+\,m^{2}_{N}\mid\!N\!\mid^{2}
-(\lambda_{N}A_{\lambda_{N}}N{H_u}{H_d}+{\frac{kA_{k}}{3}}{N^3}+h.c.),\\
&& \Delta V\ =\ {\frac{3}{32\pi^{2}}}\left[m^{4}_{\tilde
{t}_{1}}(\ln{\frac{m^{2}_{\tilde{t}_{1}}}{Q^{2}}}-\frac{3}{2})
+m^{4}_{\tilde{t}_{2}}(\ln{\frac{m^{2}_{\tilde{t}_{2}}}{Q^{2}}}
-\frac{3}{2})-2m^{4}_{t}(\ln{\frac{m^{2}_{t}}{Q^{2}}}-\frac{3}{2})\right],
\end{eqnarray}
where $V_{tadpole}$ arises from the 
$N\overline\Phi\Phi$ coupling via one-loop tadpole 
diagrams involving the messenger fields \cite{Murayama}.
For the one-loop potential term $\Delta V$, we have only included the dominant 
contributions from the top-quark and top-squark diagrams. $Q$ is
the renormalization scale in the $\overline{DR}$ scheme.
\newpage
\section{Analysis of the Model}

\subsection{Numerical Procedure}

To study physics at the electroweak scale, we perform a one-loop 
renormalization group (RG) evolution analysis.
The relevant RGEs can be derived from \cite{rge} and are listed in Appendix A.

We fix the boundary conditions for the RG evolution as follows:
At the messenger scale, we impose the conditions in Eqs.(8)-(11).
The coupling constant $\xi_{S}$ is fixed at the messenger scale
to be unity.
$\xi_{N}, \eta_{S}$ and $\eta_{N}$ are solved at the electroweak scale 
by minimizing the potential as described below. 
Values of the other three couplings,
$\lambda_{S}, \lambda_{N}$ and $k$ are chosen randomly at the
electroweak scale. 
We require the couplings to be in the perturbative ranges
\begin{equation}
 -1  \leq  \xi_{N}(S) \leq 1,\quad
 0 \leq  \lambda_S,\ \lambda_N,\ \eta_N \leq 1,\quad
 0 \leq  k \leq 0.65,
\label{range}
\end{equation}
where the upper bound on $k$ comes from the demand 
that it not hit the Landau 
pole below the GUT scale. If we require that $k$ be perturbative
only below the messenger scale $\Lambda$ where unspecified new
physics sets in, we find $k \leq 1.36$.
It is found to be unnecessary to place bounds on the value of
$\eta_{S}$. In the numerical analysis, we will neglect Yukawa coupling
terms in the RGEs, except for those involving 
the top quark ($h_t$).
The $A$-terms run slowly and so can be approximated by their boundary values.
 All RGEs are run down to
$Q=\sqrt{m_{\tilde{t}_1}m_{\tilde{t}_2} }$.
 
In the RGE running, $m_{H_u}^2$ receives large correction from 
the heavy stop loop through the Yukawa coupling $h_t$,
\begin{equation}
 m_{H_{u}}^{2}(Q) \simeq 
m_{H_{u}}^{2}(S)-{\frac{3}{8\pi^{2}}}h_{t}^{2}
\left(m_{\tilde{t}_L}^{2}(S)+m_{\tilde{t}_R}^{2}(S)+
m_{H_{u}}^{2}(S)\right)\ln \frac{S}{Q},
\end{equation}
and becomes negative at the electroweak scale, triggering EWSB.   
A satisfactory estimate for $m_{N}^{2}$ at the electroweak scale can be 
found by approximating the RGE for $m_{N}^{2}$:
\begin{eqnarray}
  m_{N}^{2}(Q)\simeq &m_{N}^{2}(S)& -{\frac{\lambda_{N}^{2}}{4\pi^{2}}}
\left(2m_{H_{d}}^2(S)-{\frac{3}{4\pi^2}}h_t^2 m_{\tilde{t}}^{2}(S)\ln {\frac{S}
{Q}}+m_{N}^{2}(S)+A_{\lambda_N}^2\right)\ln \frac{S}{Q} \nonumber \\&  & -{\frac{k^2}{4\pi^{2}}} \left(3m_{N}^{2}(S)+A_k^2\right)\ln \frac{S}{Q} .
\end{eqnarray}
which can be either positive or negative depending on the relative values 
of the couplings.
  
The physical mass spectrum at the electroweak scale is determined
by the five VEVs
\begin{equation}
\langle H_{u}\rangle=v_u,\quad \langle H_{d}\rangle=v_d,
\quad \langle N\rangle=x,\quad  
\langle S\rangle=y\quad {\rm and}\quad F_S.
\end{equation}
The values of $M_Z$ and top-quark mass fix the scale 
\begin{equation}
v\equiv\sqrt{v_u^2+v_d^2}=174\ \gev,\quad
m_t\equiv h_t v_u=165\pm 5\ \gev 
\end{equation}
where we have used the $\overline{DR}$ value of the top-quark 
running mass.
We also require that $\tan\beta=v_u/v_d$ satisfy
\begin{equation}
 1.5  \leq  \tan\beta \leq 50,
\end{equation}
which leads to $0.92 \leq  h_{t} \leq 1.2.$
We impose the following constraints on the other VEVs
\begin{equation}
   100  \leq  x \leq 1000 \ \gev,  ~~~
    3.8\times 10^{4}  \leq  \Lambda ,\  y \leq 10^{5}\ \gev, ~~~
    \Lambda   \leq  0.9y,
\label{xy} 
\end{equation}
where $\Lambda=F_S/y$. The lower bound on $\Lambda$ arises from 
the right-handed slepton mass constraint (see below), 
and the last condition
in Eq.~(\ref{xy}) is adopted to avoid possible large corrections from
$\Lambda\approx y$. 
The calculated mass spectrum for the sparticles
must satisfy the current lower limits
from direct experimental searches \cite{TEV,LEP}
\begin{eqnarray}
&& m_{\tilde g} > 190\  \gev, ~~~m_{\tilde{e}_{R}}  >  80\  \gev,  \nonumber \\
&& m_{h} \equiv  m_{S_{1}} > 95.5\  \gev,~~~
 m_{A}\equiv m_{P_{1}}  >  84.5\  \gev,~~~ m_{H^\pm} > 77.3\  \gev,  \nonumber \\
&& m_{\tilde\chi_{1}^{\pm}}   >  95\  \gev,~~~
   m_{\tilde{b}_{1}}  >  75\  \gev,  \nonumber \\
&& {\rm if}\  {\frac{m_{\tilde\chi_1^0}}{m_{\tilde{\tau}_{{1}}}}} <1,\   
m_{\tilde\chi_1^0}  >  55\ \gev;~~~
{\rm if}\  {\frac{m_{\tilde\chi_1^0}}{m_{\tilde{\tau}_{1}}}} >1,\   
m_{\tilde{\tau}_{1}}  >  68\ \gev;  \nonumber \\
&&{\rm if}\ {\frac{m_{\tilde\chi_1^0}}{m_{\tilde{\tau}_{1}}}} >1\  
{\rm and}\ m_{\tilde\chi_1^0}<87\ \gev\  {\rm then}\  
m_{\tilde{\tau}_{1}} > 84\  \gev. 
\label{exp}
\end{eqnarray}

The minimization conditions to be imposed at ${\cal O}(M_{Z})$ 
can be derived by differentiating Eq.~(\ref{pot}) with respect to $v_d$, $v_u$
and $x$:
\begin{eqnarray}
&&2 m_{H_{d}}^{2} {v_d} + 
{\frac{\bar{g}^{2} {v_d} \left( v_{d}^{2} - v_{u}^{2} \right) }{2}} +  {v_u} 
{{\lambda }_N} \left( -2 k {x^2} - 2 x y {{\eta }_N} - {y^2} {{\eta }_S} + 2 
{v_d} {v_u} \lambda_N \right) +  2 {v_d} {{\left( x {{\lambda }_N}
  + y {{\lambda }_S} \right) }^2}
\nonumber \\ & &\qquad
 + 
  {v_u} {{\lambda }_S} \left( 2 {F_S} - {x^2} {{\eta }_N} - 
2 x y {{\eta }_S} + 2{v_d}{v_u}{{\lambda }_S} \right) 
+ \partial_{v_{d}} \Delta V = 0 ,
\label{min1}
\\
&&2 m_{H_{u}}^{2} {v_u} - 
{\frac{\bar{g}^{2}  {v_u} \left( v_{d}^{2} - v_{u}^{2}  \right) }{2}} 
+   {v_d} 
{{\lambda }_N} \left( -2 k {x^2} - 2 x y {{\eta }_N} - {y^2} {{\eta }_S} + 2 
{v_d} {v_u} {{\lambda }_N} \right) +  2 {v_u} {{\left( x {{\lambda }_N}
 + y {{\lambda }_S} \right) }^2} 
\nonumber \\ & &\qquad
+ {v_d} {{\lambda }_S} \left( 2 {F_S} - {x^2} {{\eta }_N} - 
2 x y {{\eta }_S} + 2 {v_d} {v_u} {{\lambda }_S} \right) 
+ \partial_{v_{u}} \Delta V=0, \label{min2}
\\
&&2 m_{N}^{2} {x}\  + {\frac{F_{S}^{2}{{\xi 
}_N}{{\xi }_S}}{8{{\pi }^2}y}} + \left( 2kx + y{{\eta }_N} \right) \left( 
2k{x^2} + 2xy{{\eta }_N} + {y^2}{{\eta }_S} - 
 2{v_d}{v_u}{{\lambda }_N} \right) 
+ 2 v^{2} {{\lambda }_N}\left( x{{\lambda }_N} + y{{\lambda }_S} \right)
\nonumber \\ & &\qquad
+ \left( x{{\eta }_N} + y{{\eta }_S} \right) \left( -2{F_S} 
+ {x^2}{{\eta }_N} + 2xy{{\eta }_S} - 
 2{v_d}{v_u}{{\lambda }_S} \right) + \partial_{x} \Delta V =0,
\label{min3}
\end{eqnarray}
where $\bar{g}^2={g'}^2+g^{2}_{2}$.  
Running $\xi_{S}$ down to the electroweak 
scale, we get $\xi_{S}(Q) \simeq 0.94$.
$\eta_{S}, \eta_{N}$ and $\xi_{N}$ are determined from 
the minimization conditions Eqs.~(\ref{min1})-(\ref{min3}). 
These equations also lead to the relations
\begin{eqnarray}
\mu^{2}&=&-{\frac{M_{Z}^{2}}{2}}
+{\frac{[m_{H_{d}}^{2}+ (\partial_{v_{d}} \Delta 
V)/(2 v_{d})]-[m_{H_{u}}^{2}+(\partial_{v_{u}} \Delta V)/(2 v_{u})] 
\tan^{2}\beta}{\tan^{2}\beta-1}}, \\
\sin2\beta&=& {\frac{2B_{\mu}}{[m_{H_{d}}^{2}+ (\partial_{v_{d}} \Delta V)/(2 
v_{d})]+[m_{H_{u}}^{2}+(\partial_{v_{u}} \Delta V)/(2 v_{u})]+2\mu^{2}}},
\label{Bmu0}
\end{eqnarray}
where
\begin{eqnarray}
 \mu &\equiv& \lambda_{N} x +\lambda_{S} y, \\
B_{\mu}&\equiv& \lambda_{N}(A_{\lambda_N} x+kx^{2}+\eta_{N} xy
+{\frac{\eta_{S} y^{2}}{2}}-\lambda_{N} 
v_{u}v_{d})+ \lambda_{S}(-F_{S}+\eta_{S} xy+{\frac{\eta_{N} 
x^{2}}{2}}-\lambda_{S} v_{u}v_{d}).
\label{Bmu}
\end{eqnarray}
The parameters $\mu$ and $B_{\mu}$ can be evaluated by either set of
 equations and they must be consistent with each other since all 
the parameters have been determined by conditions on EWSB.

\subsection{Viable solutions, $\mu$ and $B_{\mu}$ parameters}

\begin{figure}[thb]
%\centerline{
\mbox{\psfig{file=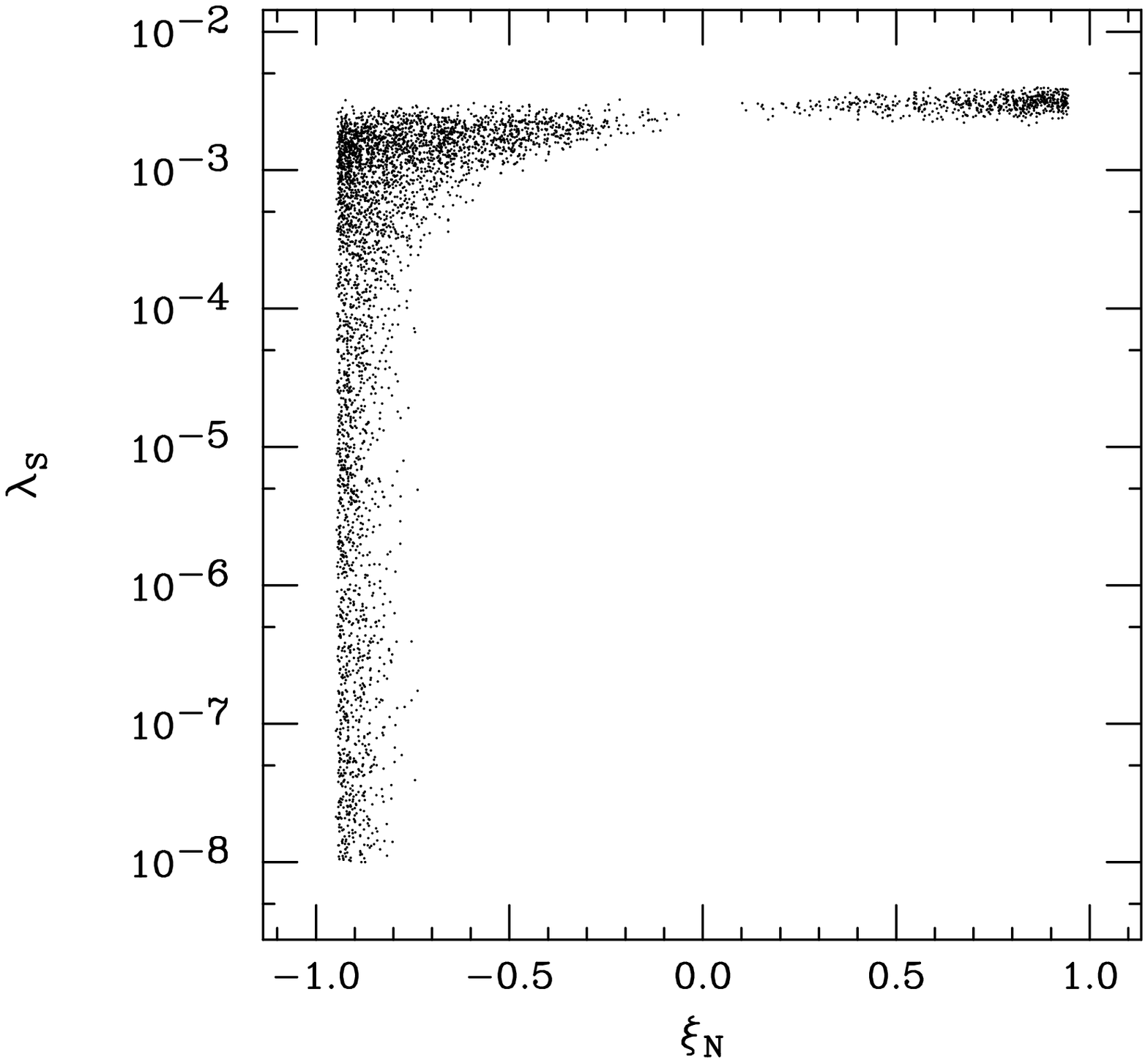,angle=90,width=8.7cm,height=8cm}
\psfig{file=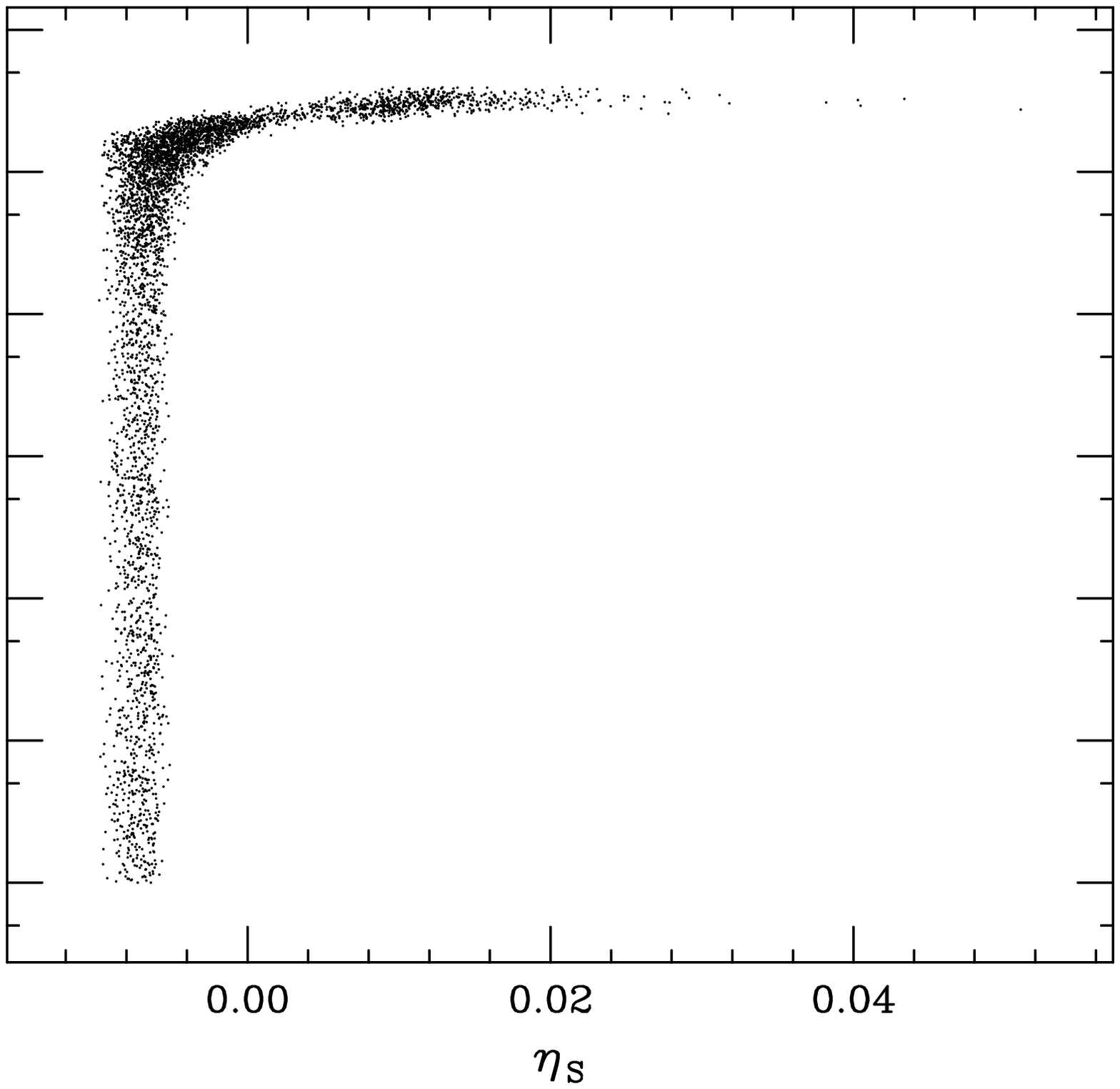,angle=90,width=6.75cm,height=8cm}}
\bigskip
\mbox{\psfig{file=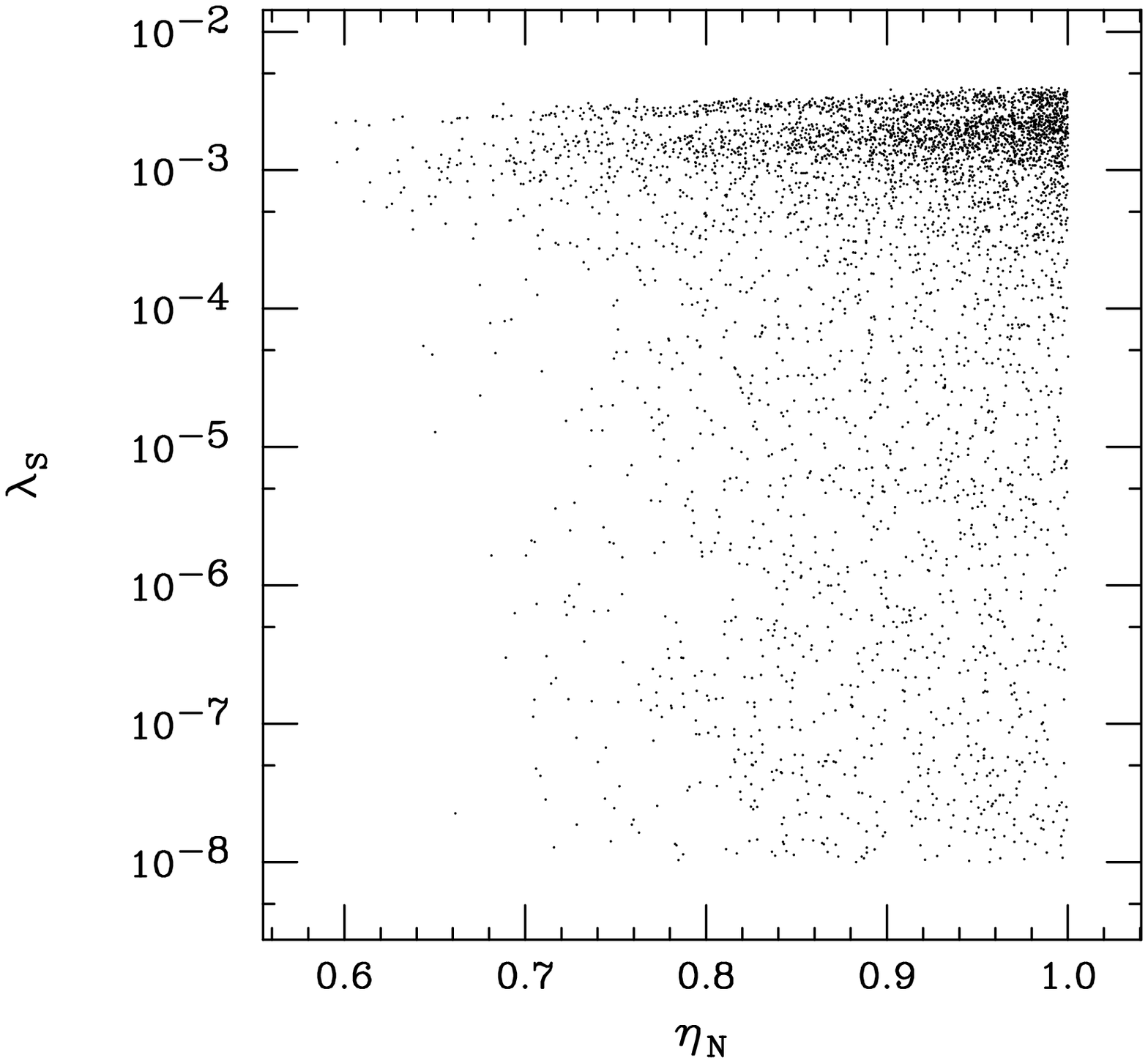,angle=90,width=8.75cm,height=8cm}
\psfig{file=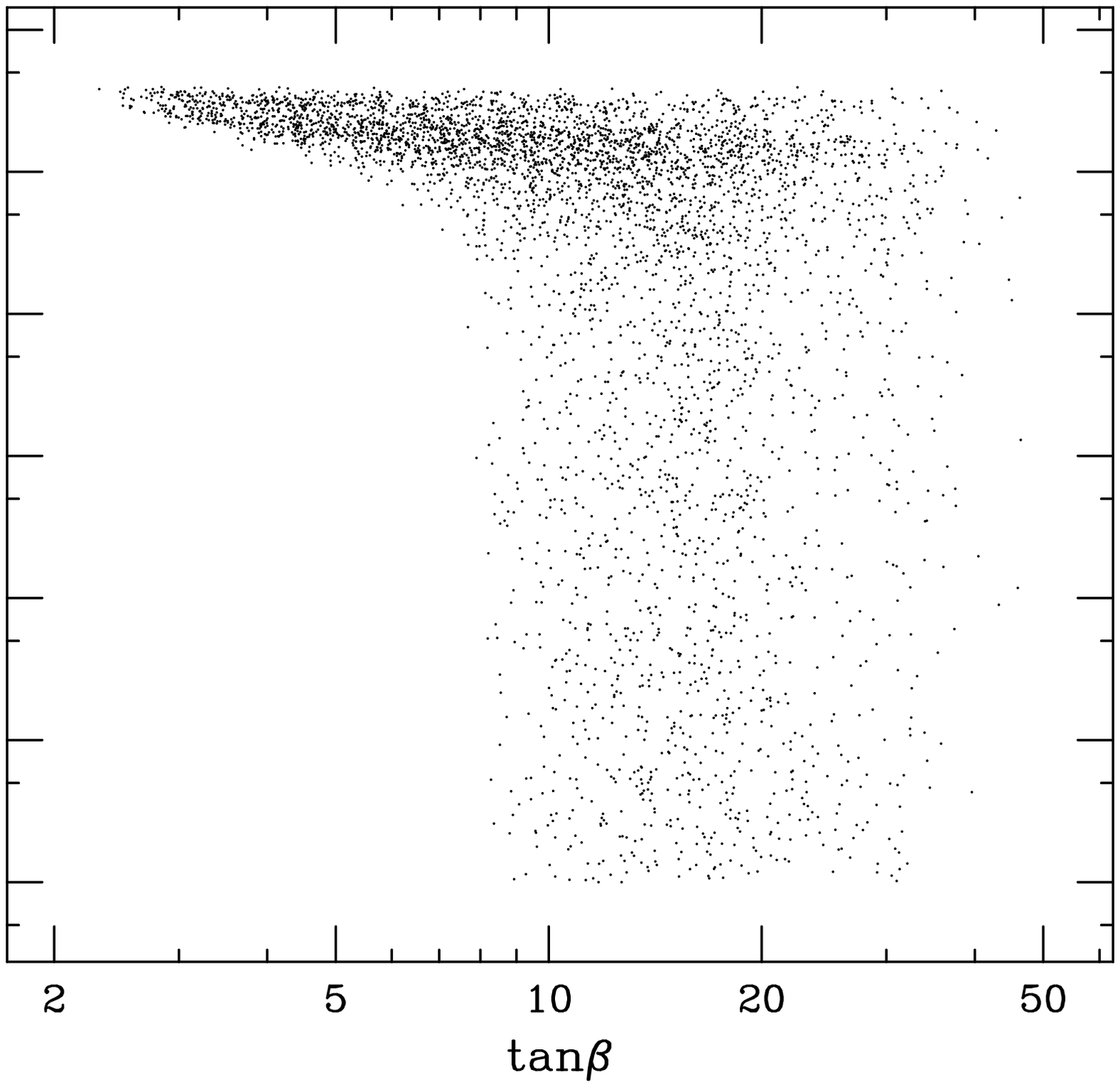,angle=90,width=6.75cm,height=8cm}}
\bigskip
\caption[]{Phenomenologically viable solutions in the $\lambda_{S}$--$\xi_N^{}$,
 $\lambda_{S}$--$\eta_S^{}$, $\lambda_{S}$--$\eta_N^{}$ and 
$\lambda_{S}$--$\tan\beta$    planes.}
\label{lambdaSetaS}
\end{figure}

On inspection of the parameter space of the variables chosen randomly in the 
allowed ranges in Eq.~(\ref{range}), it is found that there are solutions 
satisfying the theoretical and experimental constraints of 
Eqs.~(\ref{xy})-(\ref{exp}).
We first note that $\eta_N$ and the ``NMSSM couplings'' $\lambda_{N}$ 
and $k$ are essentially independent of other parameters and mostly of
 ${\cal O}(1)$,
\begin{equation}
0.6\alt  \eta_{N} \alt 1.0,\quad 
0.05\alt  \lambda_{N} \alt 1.0,\quad 0.001  \alt \  k\  \alt 0.65 .
\end{equation} 
 $\xi_N$ and $\eta_S$ depend quite crucially on the values of $\lambda_S$
and take on values of ${\cal O}(1)$ and \mbox{${\cal O}(10^{-3}-10^{-2})$}
 respectively,
\begin{equation}
-0.95  \alt  \xi_{N} \alt 0.95\ {\rm and}\  \mid\!\xi_{N}\!\mid\, \agt 0.1, 
\end{equation}
\begin{equation}
   \ \    -0.01  \alt  \eta_{S} \alt 0.05\ {\rm and}\  \mid\!\eta_{S}\!\mid\, \agt 10^{-4},
\end{equation}
corresponding to
\begin{equation}
   0  \alt \lambda_{S} \alt 4 \times 10^{-3}.
\end{equation}

%Note that it is possible to find solutions 
%with $\lambda_{S}\not\simeq 0$.
The correlations for those parameters are shown in 
Fig.~\ref{lambdaSetaS}. There are two rather distinctive regions
in $\lambda_{S}$.
\begin{enumerate}
\item Vanishing $\lambda_{S}$: 
There are viable solutions
for vanishing $\lambda_{S}$. In this case, typically
$\lambda_{S} \alt 10^{-3}$, leading to tightly 
constrained solutions for the allowed values of $\xi_{N}$ 
\mbox{and $\eta_{S}$}.
\begin{equation}
 -0.95  \alt \xi_{N} \alt -0.3 ,\ \
 -0.01  \alt \eta_{S} \alt -4 \times 10^{-3}.
\end{equation} 
It is interesting to note that $\eta_{S}$ and $\xi_{N}$ are always negative, 
 thus picking out a particular choice of phases in the superpotential
\begin{displaymath}
 W \supset \xi_{S}S\Phi\overline\Phi\,-\mid\!\xi_{N}\!\mid 
N\Phi\overline\Phi+{\frac{\mid\!\eta_{S}\!\mid}{2}}{S^2}N+\ldots    
\end{displaymath}
Also notice that small $\tan\beta$ is excluded. 
\item Finite  $\lambda_{S}$: 
For $\lambda_{S} \agt 10^{-3}$, the couplings $\xi_{N}$ and  
$\eta_{S}$ can take on the whole range of permissible values.
 Moreover, small $\tan\beta$ solutions are allowed.
\end{enumerate}
Since we wish to keep all terms in the superpotential that
 respect the $\mathbf{Z_3}$ symmetry, we avoid 
a vanishing $\lambda_{S}$ and restrict our attention to the case
$\lambda_{S} \geq 10^{-3}$ in the rest of the paper. This choice gives
us greater freedom in parameter space.

\begin{figure}[p]
\centerline{\psfig{file=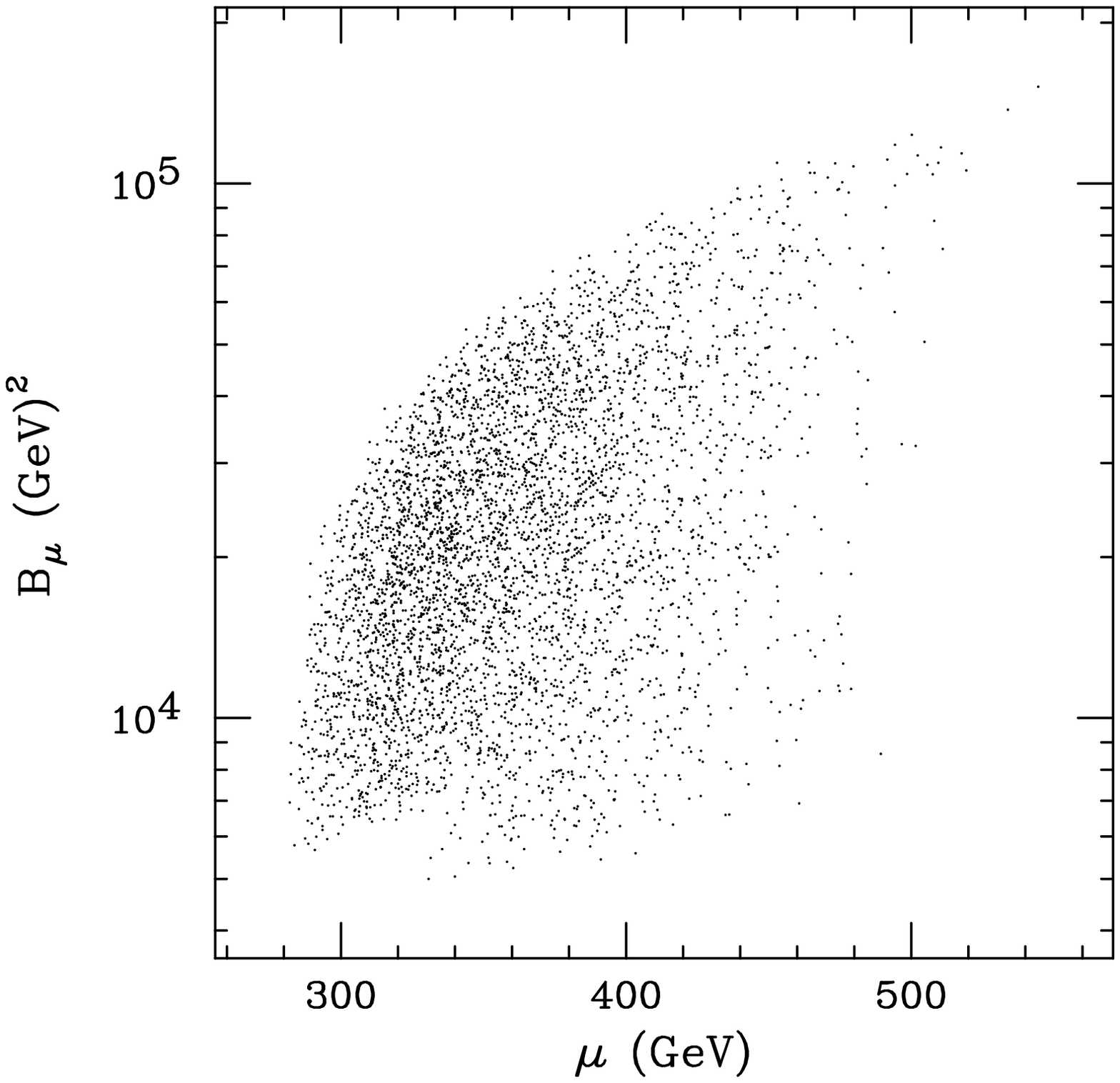,angle=90,width=8cm,height=8cm}
\psfig{file=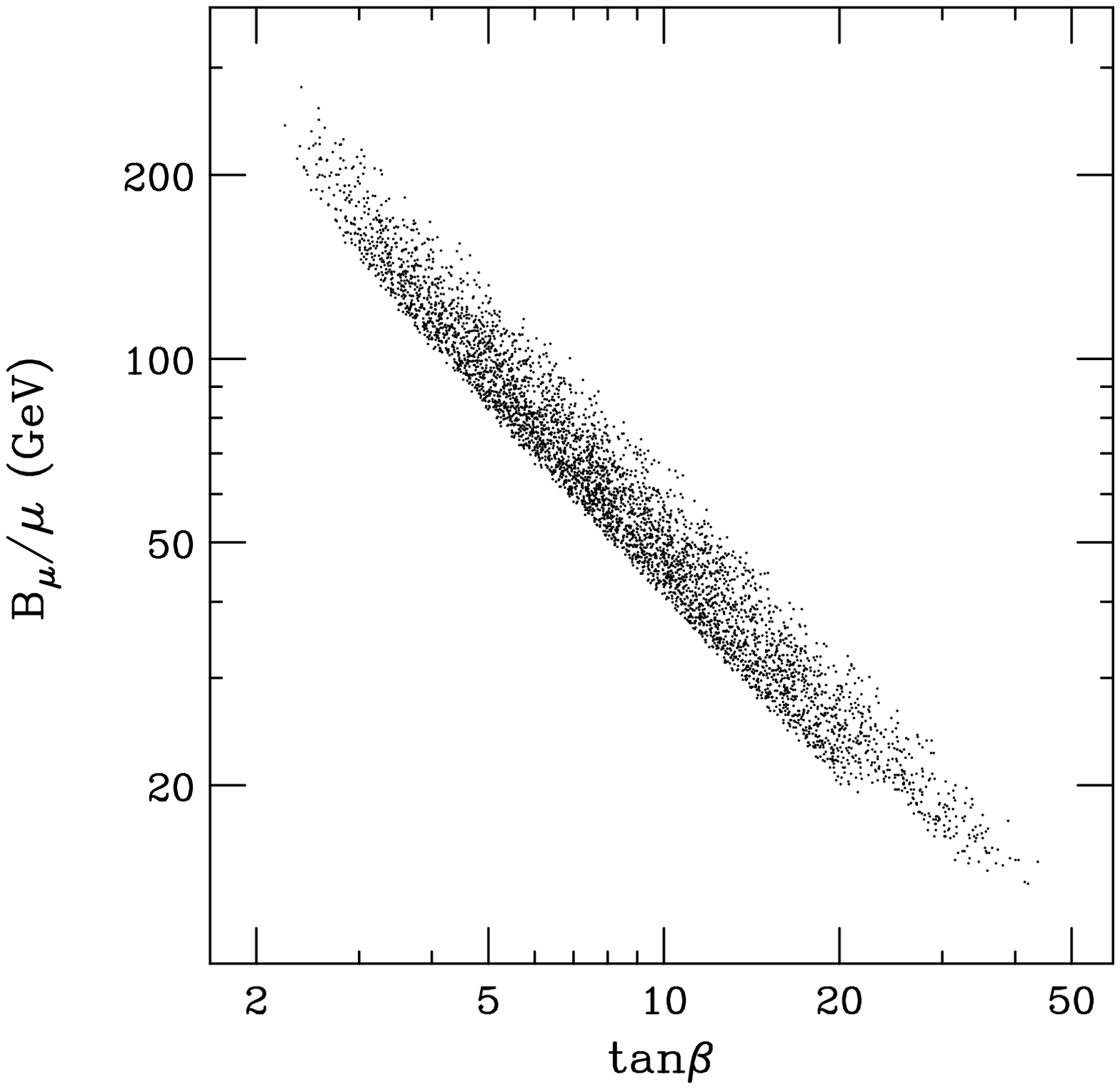,angle=90,width=8cm,height=8cm}}
\bigskip
\caption[]{The allowed region in the (a) $B_{\mu}$--$\mu$ plane
and (b) ($B_\mu/\mu$)--$\tan\beta$ plane. }
\label{Bmumu}
\end{figure}

\begin{figure}[p]
\centerline{\psfig{file=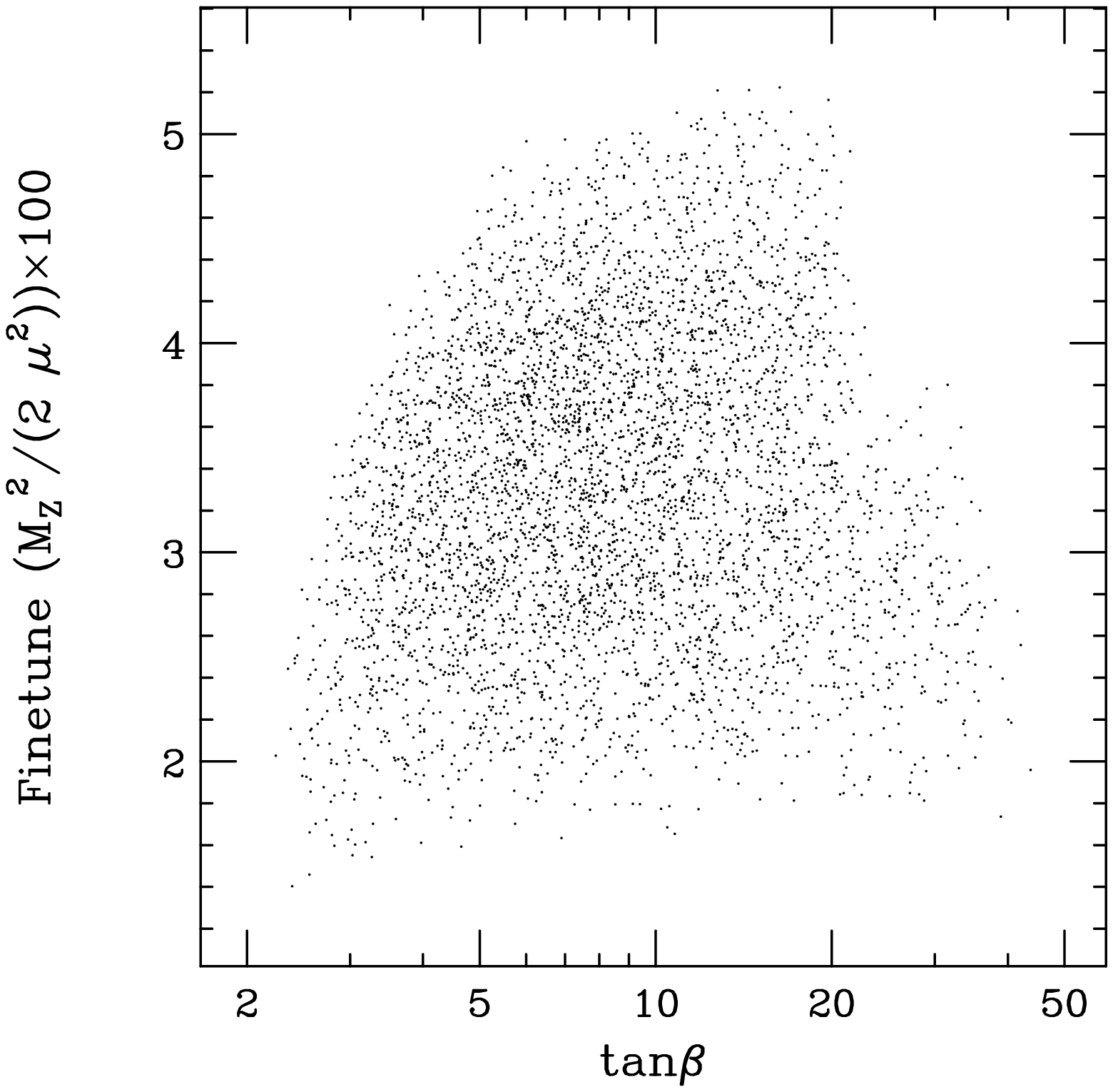,angle=90,width=8cm,height=8cm}
\psfig{file=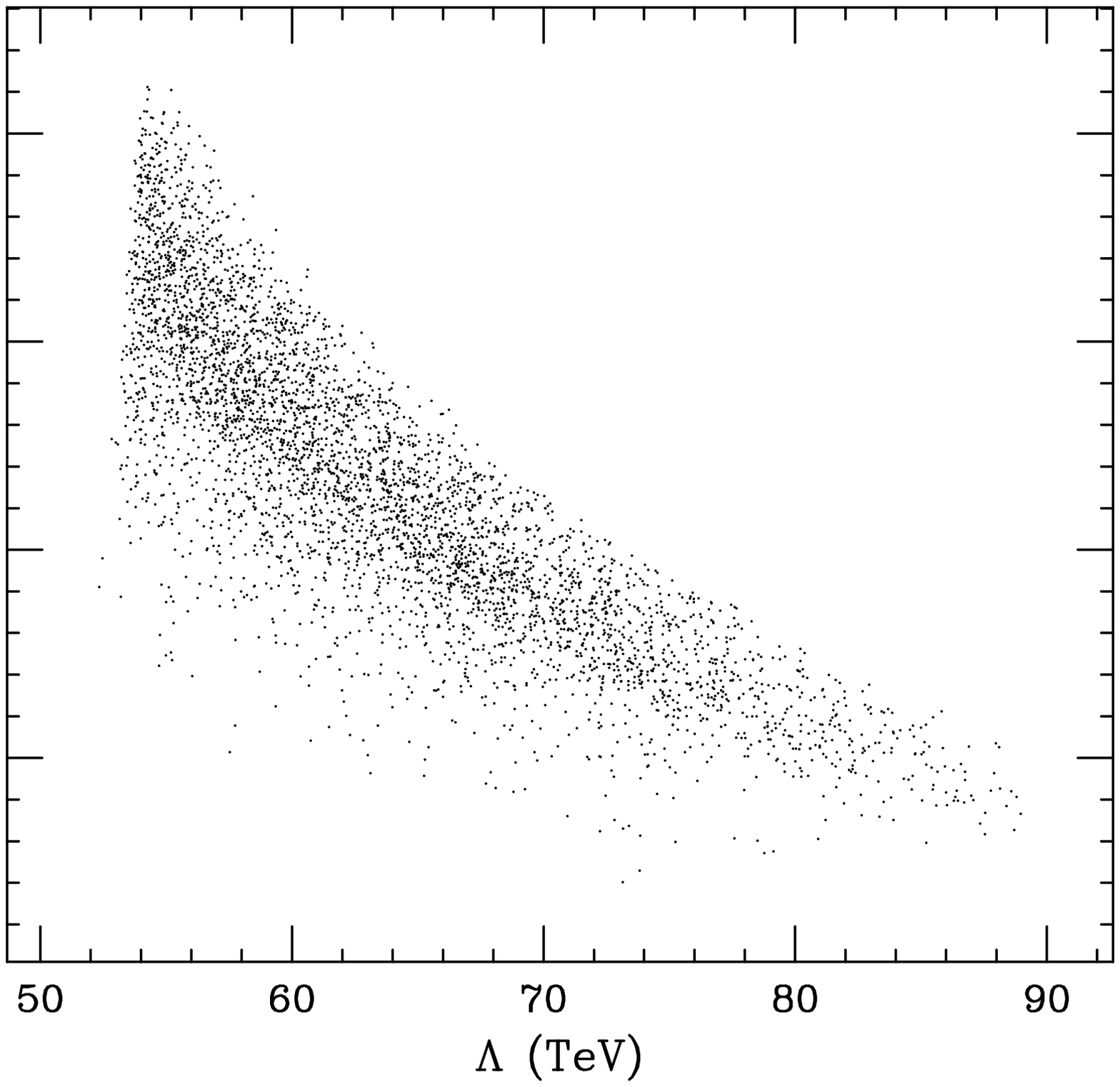,angle=90,width=6.6cm,height=8cm}}
\bigskip
\caption[]{The percentage fine-tune versus
$\tan\beta$ and $\Lambda$.}
\label{finetunetanbeta}
\end{figure}

The parameters $\mu$ and $B_{\mu}$ are spontaneously generated 
and are of ${\cal O}(M_Z)$ and ${\cal O}(M_Z^2)$, respectively. 
As shown in Fig.~\ref{Bmumu}(a),
the values lie in the ranges 
\begin{equation}
275 \alt \  \mu \  \alt 550\  \gev ,\quad
 5 \times 10^{3}  \alt  B_{\mu}  \alt 2 \times 10^{5}\  (\gev)^{2} .
\end{equation} 
Moreover the ``$B$-parameter'' ($B_{\mu} / \mu$)
familiar from supergravity theories is also ${\cal O}(M_{Z})$ 
for most of the parameter space 
\begin{equation}
 10  \alt  {\frac{B_{\mu}}{\mu}}  \alt 300\ \gev,
\end{equation} 
as shown in Fig.~\ref{Bmumu}(b). 
% in Fig.~\ref{Bmumutanbeta}. 
For $\tan\beta \agt 2$, $M_Z^2/2 \simeq -\mu^{2}-m_{H_{u}}^{2}$.
A good measure of fine-tuning of the Higgs potential parameters is 
$M_Z^2/2\mu^2$ \cite{finet}. 
The fine-tune is at the 1-6\% level as seen in
Fig.~\ref{finetunetanbeta}, where the fine-tune parameter is shown
versus $\tan\beta$ and $\Lambda$. This level of fine-tune could 
be significantly improved if one introduces extra vector-like triplets
that couple only to the singlet $N$ \cite{finetune}.
Recall that in the MSSM, $\mu$ of electroweak scale is put in by hand.
Even if the very lowest value of $\mu$ (consistent with 
$ m_{\tilde{e}_{R}} \agt 80$ GeV) is used, the best possible 
fine-tune is 7\%.


\subsection{Mass Spectrum}

The details of sparticle spectroscopy of models with gauge mediation have been 
discussed in the literature over quite large ranges 
of both $\tan \beta$ and $\Lambda$ \cite{gmsbpheno}. 
Many of those results carry-over unchanged to the case at hand. 
As one would expect, the main differences arise from the specific 
choice of the Higgs sector superpotential. 
The mass-squared matrices for the CP-even and CP-odd Higgs bosons and 
their eigenvalues are presented in the Appendix B. 

We find the range for the CP-even Higgs boson masses
\begin{equation}
95.5 \alt m_{S_{1}} \alt 130\ \gev ,\qquad
200  \alt  m_{S_{2}} \alt 650\ \gev ,\qquad
 m_{S_{3}} \agt 5\ \tev.
\end{equation}
We see that $m_{S_1}$ has the usual upper bound of 130 GeV. 
If the perturbativity for $k$ is relaxed to the messenger scale,
 $m_{S_1}$ can be as heavy as about 170 GeV.
In the NMSSM, there are two CP-odd Higgs boson mass eigenstates 
($P_1,P_2$). 
 The mass of the lighter CP-odd state lies in the range
 $340 \alt  m_{P_{1}} \alt 650$ GeV indicating that we are well into the 
decoupling regime ~\cite{decoupling}.
$P_1$ is primarily a mixture of Im$H_d$ and Im$H_u$ since  the mixing angle 
$\gamma$ between $A^0$ 
and Im$N$ is ${\cal O}({10}^{-4})$ (see Appendix B). We also find
that the charged Higgs bosons $H^{\pm}$ are almost degenerate with $P_1$.
%From Fig.~\ref{mS1tanbeta}, we see that 
%when the mixing is small, $\sin\gamma \rightarrow 0$, 
%$m_{P_{1}}$ tends to be above one TeV.
%
%\begin{figure}[p]
%\centerline{\psfig{file=Fig4.ps,angle=90,width=8cm}}
%\bigskip
%\caption[]{Allowed masses of the lighter CP-odd state in the 
%$m_{P_1}$--$\sin\gamma$ plane.}
%\label{mS1tanbeta}\end{figure}
%

\begin{figure}[th]
\centerline{\psfig{file=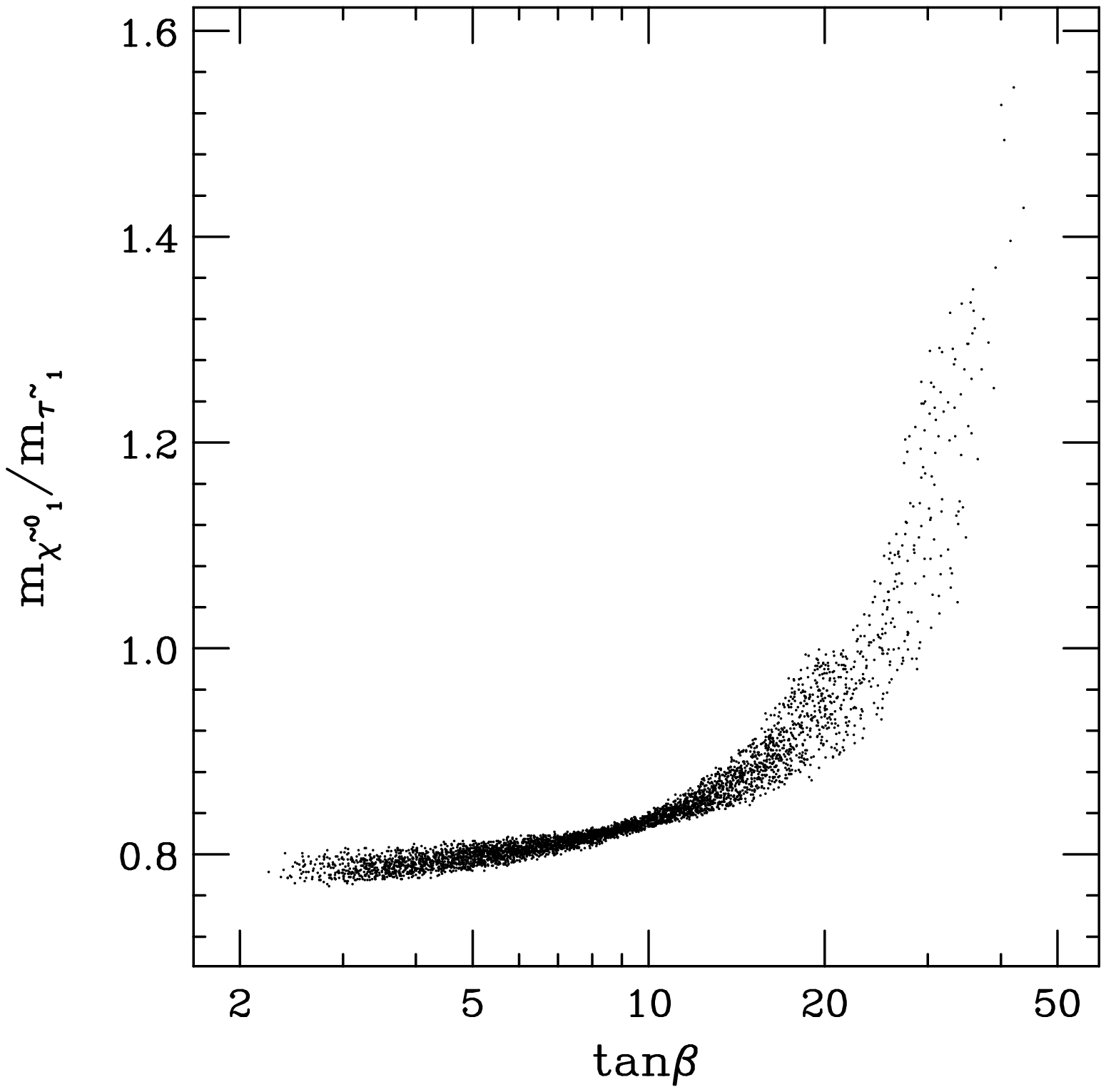,angle=90,width=8cm,height=8cm}
\psfig{file=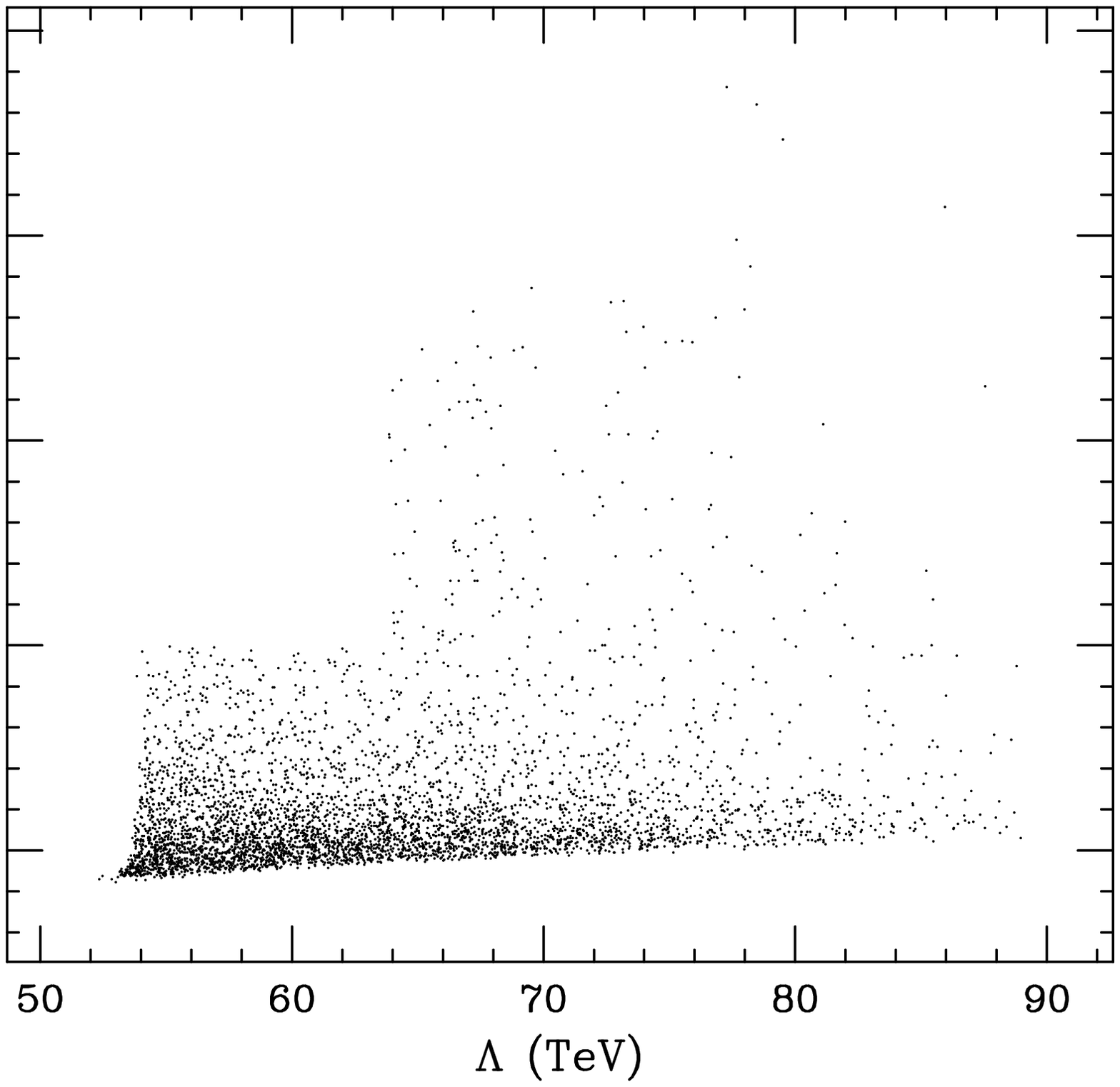,angle=90,width=6.6cm,height=8cm}}
\bigskip
\caption[]{Regions of $\protect\tilde{\chi}_1^0$-NLSP and
$\protect\tilde{\tau}_1$-NLSP in the 
($m_{\tilde{\chi}_{1}^{0}}/m_{\tilde{\tau}_{1}}$)--$\tan\beta$ and 
($m_{\tilde{\chi}_{1}^{0}}/m_{\tilde{\tau}_{1}}$)--$\Lambda$ planes.}
\label{NLSPtanbeta}
\end{figure}

Since $\mu$ is large compared to $M_{Z}$, the higgsinos and gauginos 
decouple from each other and the lightest neutralino is mostly bino.
It is the $\tilde{\chi}_{1}^{0}$-NLSP for a large region of parameter space 
\begin{equation}
\tan\beta  \alt  22\quad {\rm or}\quad \Lambda  \alt  64\ \tev.
\end{equation}
Due to the mixing in the $\tau$-slepton (stau) mass matrix, 
$m_{\tilde{\tau}_{1}}$ can be much smaller than 
$m_{\tilde{e}_{R},\tilde{\mu}_{R}}$ 
and becomes the NLSP for large $\tan\beta$ and high values of $\Lambda$.
These features are shown in Fig.~\ref{NLSPtanbeta}.
We find the mass ranges for the lighter sparticles are
\begin{equation}
65  \alt  m_{\tilde{\chi}_{1}^{0}} \alt 125\ \gev,\quad
120  \alt  m_{\tilde{\chi}_{1}^+} \alt 230\ \gev,\quad
68  \alt  m_{\tilde{\tau}_{1}}  \alt 150 \ \gev.
\end{equation}

Gluinos and squarks are much heavier than the electroweak 
states because they 
acquire mass mainly from strong interactions. We find the ranges the of
stop masses as follows:
\begin{equation}
650 \alt  m_{\tilde{t}_{1}} \alt 1080 \ \gev, \qquad
700 \alt  m_{\tilde{t}_{2}} \alt 1140 \ \gev.
\end{equation}
Table~\ref{spectrum} gives a representative set of
parameters and the corresponding mass spectrum.

\begin{table}[thb]
\begin{center}
\begin{tabular}{|c|c|c|c|c|c|c|c|c|c|c|}
$\tan \beta$ & $\xi_S$ & $\xi_N$ & $\eta_S$ & $\eta_N$ & $\lambda_S$ &
 $\lambda_N$ & $k$ & $\langle S\rangle$ & $\Lambda$ & 
$\langle N\rangle$  \\
\hline
24.3 & 0.94   & 0.71 &$ 8.3 \times 10^{-3}$&
$0.87$&$3.0 \times 10^{-3}$& 
0.31 & 0.40 & 97 TeV & 77 TeV & 413
\end{tabular}
\begin{tabular}{|c|c|c|c|c|c|c|c|c|c|c|c|c|c|}
$\mu$ &$M_1$ & $M_2$ & $M_3$ & $m_{\tilde\chi_1^0}$ & $m_{\tilde\chi_1^+} $ &
$m_{S_1}$ & $m_{P_1}$& $m_{H^\pm}$ & $m_{\tilde \tau_1} $& $ m_{\tilde \tau_2} $ & 
$m_{\tilde t_1} $  & $m_{\tilde t_2} $ & $m_{\phi^0}$ \\
\hline
421 & 107 & 208 & 649 & 106 & 196 & 122 & 500 & 505  & 109 & 288 & 949 & 982 
& 10.7 TeV
\end{tabular}
\parbox{6.0in}{
\caption[]{\small A representative set of parameters and 
the mass spectrum. All masses are in GeV unless specified.}
\label{spectrum} }
\end{center}
\end{table}

\section{Messenger Sneutrinos as Cold Dark Matter}

The possibility of messenger sneutrinos ($\phi^0$) 
as a cold dark matter (CDM) candidate
 was investigated in the context of the minimal GMSB model in Refs.
\cite{messenger-dm,cdm}. Requiring that the Universe not be 
over-closed by the messenger sneutrinos ($\Omega_{\phi^0} h^2 < 1$) 
puts an upper bound on their masses
\begin{equation}
m_{ \phi^0} < 3\  \tev.
\end{equation}
It certainly requires some degree of fine-tuning to obtain such a 
light messenger particle as the messenger scale is about $50-100$ TeV.
More severely, a scalar DM particle with mass lighter than 3 TeV 
and with SM interactions has already been experimentally ruled 
out \cite{sneu}.
In the GMSB model with an extra singlet Higgs field, 
additional contributions from the singlet $N$ can sufficiently increase
the annihilation cross-section of $\phi^0$, and therefore 
loosen the above bound. 
\begin{figure}[th]
\centerline{\psfig{file=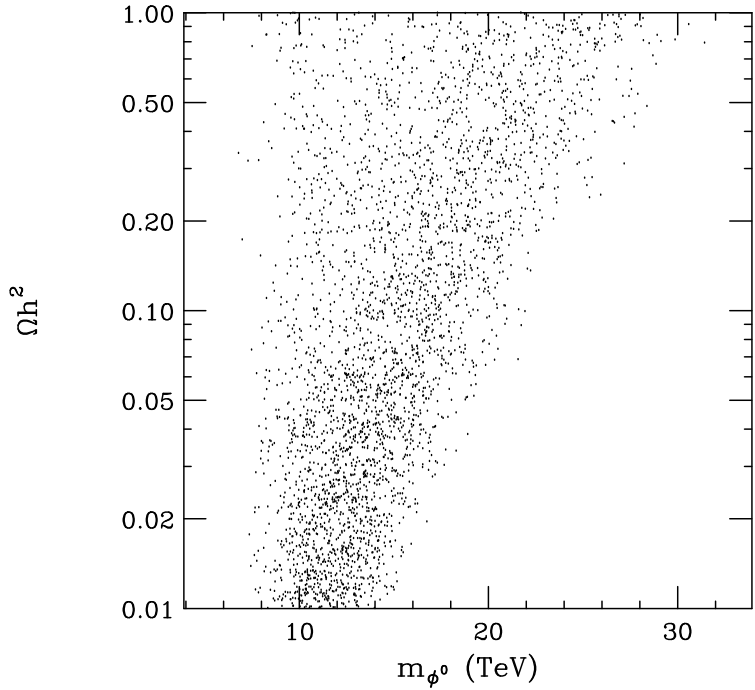,angle=90,width=10cm}}
\bigskip
\caption[]{Messenger sneutrino masses that do not over-close the universe.}
\label{relic}
\end{figure} 

We perform a relic density calculation in our model. We first note 
that the freeze-out temperature \mbox{ $T_f \simeq m_{\phi^0}/20\gg 
m_{\phi^-} -m_{\phi^0}$}, which implies that both charged and neutral
components are present in the thermal bath at $T_f$ \cite{cdm}.
We calculate the relic density in
the limit of symmetric phase for the four components 
($\phi^Q, \phi^{Q\dagger}$ with $Q=0,-$). On examining the  
interaction  vertices, by far the dominant annihilation processes are 
\begin{equation}
 \phi \phi^\dagger \rightarrow N N, H_u H_d
\end{equation}
via the $t$- and $u$-channel messenger exchange and $s$-channel
singlet exchange. 
%where the singlet 
%Higgs field has been decomposed into its real and imaginary
%components $N=(N_R+iN_I)/\sqrt 2$. 
Following \cite{Relic}, 
we define a Lorentz invariant function 
\begin{equation}
w(s) \equiv {1 \over 64 \pi} \int_{-1}^{1}
{1\over 8}\sum_{Q, Q^\prime,a,b}
|{ {\cal M}(\phi^{Q\dagger} \phi^{Q^\prime} \rightarrow a\, b)}|^2
 d\cos\theta
\end{equation}
where $\sqrt s$ is the center of mass energy. Specifically
in our model,
\begin{equation}
% w(s)= \frac{(\xi_N \xi_S r)^4}{64 \pi} \frac{ m_{\phi}^2}{s}
 w(s)=\left({\frac{(\eta_N k \vev{S})^2 \vartheta}{16 \sqrt{\pi} s}}
\right)^2 \left[1-{\frac{1}{\sqrt{2}}}+2({\frac{\lambda_N}{k}})^2-
{\frac{4}{3}}\vartheta\left(1-{\frac{m_{\phi}^2}{s}}\right)
+\vartheta^2{\frac{s}{4 m_{\phi}^2}}\right] 
\end{equation}
%where $r = \vev{S}/m_{\phi^0}$.
where $\vartheta=2\xi_N(2\xi_S-\eta_N)/(\eta_N k)$. 
With $\kappa=T_f/m_{\phi^0}$ and 
\begin{equation}
A = w(4 m_{\phi^0}^2), \quad 
B = 3 \left[2 m_{ \phi^0}^2\ {\partial w(s)\over \partial s} - 
w(s)\right]_{s = 4 m_{ \phi^0}^2}
%=\frac{9(\xi_N \xi_S r)^4}{512\pi},
\end{equation}
the thermal average 
$\vev{{\overline \sigma} 
v_{\rm rel}}$ can be expressed as
\begin{equation}
\vev{{\overline \sigma} v_{\rm rel}} =
{1\over m_{ \phi^0}^2} (A + B\kappa)
\end{equation}
It is customary to express the relic abundance in terms of 
the mass density in units of the critical density 
$\Omega_{\phi^0}=\rho_{\phi^0}/\rho_c$. It is found that
\begin{equation}
\Omega_{\phi^0} h^2 =  {8.5 \times 10^{-5} \over \sqrt{g_*} }
\ \left({m_{\phi^0}\over {\rm TeV}}\right)^2
\ {\kappa^{-1} \over A+{1\over 2}B \kappa} \ ,
\end{equation}
where $g_*\simeq 232.5$ for the particle content of the NMSSM
at the freeze-out temperature $T_f$.
The value of $\kappa$ is related to $T_f$ and is obtained iteratively by
\begin{equation}
\kappa^{-1} \equiv  m_{\phi^0}/ T_f 
= \ln \left[{0.076\over \sqrt{g_*}} 
{M_{Pl}\over m_{\phi^0}} ( A+B \kappa)\sqrt{\kappa} \right] .
\end{equation}

From the parameter space that provided a phenomenologically 
viable solution to the $\mu$-problem we select the subspace 
that does not over-close the Universe.  
Fig.~\ref{relic} shows the relic density versus the cold
dark matter mass $m_{\phi^0}$,
where $m_{\phi^0}/\vev{S}$ was allowed to vary between 0.1 and 1. 
If we assume that  $\Omega_{\phi^0} h^2 \sim 0.3$, then the mass of the 
lightest messenger sneutrino lies in the range 
\begin{eqnarray}
 6 & \alt & m_{\phi^0} \alt  25 \  \tev.
\end{eqnarray}
The smaller values of $m_{\phi^0}$ correspond to $m_{\phi^0}/\vev{S}$ 
close to 0.1 and are undesirable from the fine-tuning argument. 
We conclude that massive scalar messenger cold dark matter 
is natural in this model.  

\section{A Resolution of the Cosmological Domain Wall Problem}

It is well-known that the NMSSM predicts the formation of domain walls at the 
weak scale arising  from the spontaneous breaking of the $\mathbf{Z_3}$ 
symmetry of the superpotential. Although there are proposals 
to resolve this problem \cite{walls},
the most straightforward solution still
involves the introduction of 
higher dimensional non-renormalizable operators that 
explicitly break the discrete symmetry. It has 
been shown that this solution is unsatisfactory when SUSY-breaking
is mediated by gravity \cite{Domain}. 
We briefly indicate where the problem lies.

A wall network propagates once it has formed. The most important 
force acting on 
the domain walls that can dissolve them is pressure. Pressure arises from an 
explicit violation of the discrete symmetry. If the vacua are separated by an 
energy density $\epsilon$, the pressure is typically of order $\epsilon$. 
When this difference is 
greater than the surface tension $\sigma/R$, where $R$ is the 
curvature scale of the walls and $\sigma$ is the surface energy 
density, there is a minimum value of 
$\epsilon$ above which the dynamics will have been dominated 
by pressure before today. Thus, as the Universe expands, pressure 
catches up with surface tension, ultimately dominates it, 
the wall network is destroyed and the true vacuum is selected.

In general if the $\mathbf{Z_3}$ symmetry is broken at an energy 
scale $v_0$, then $\sigma \sim {v_0}^3$,
 and if pressure is to be dominant at an epoch  $T_*$ before the
 quark-hadron phase  transition, the curvature
 scale will be roughly the scale factor 
$R \sim M_{Pl}/T^2$, and the relation
\begin{equation}
\epsilon \agt \frac{{v_0}^3 {T_*}^2}{M_{Pl}}
\end{equation}
must hold. With $\phi$ a real scalar field having a VEV $v_0$, the 
leading higher dimensional $\mathbf{Z_3}$-breaking operator 
can be written as $\lambda'\phi^5/M_{Pl}$. 
Assuming other couplings to be ${\cal O}(1)$, the above constraint
 can now be rephrased as a constraint on $\lambda'$:
\begin{equation}
\lambda' \agt \left(\frac{T_*}{v_0}\right)^2.
\end{equation}
Note that the above relation holds only when all the fields in
 the $\mathbf{Z_3}$-breaking operator acquire the same VEV as the scale
at which the symmetry is broken. The relevance of this point will 
become clear below.

Nucleosynthesis dictates that the walls decay before the Universe 
cools to 0.1 to 1 MeV. If this were not the case, 
the abundances of the lighter elements 
would be significantly smaller than measured. 
With $T_* \sim 10$ MeV and $v_0=v\simeq 174$ GeV,
\begin{equation}
\lambda' \agt 10^{-8}.
\label{lam}
\end{equation}
However, when the mediation of SUSY-breaking is gravitational, 
this leads to destabilization of the hierarchy \cite{Domain,singlet}. 
For example, the operator $\lambda' (H_{u} H_{d}) N^{2}/M_{Pl}$ along with
$\lambda_N H_{u} H_{d} N$ generates a tadpole that leads to a term
in the superpotential of the form $\mu' N$ with
\begin{equation}
%\mu' \sim {\frac{\lambda' \lambda_{N}}{(16 \pi^{2})^{2}}} m_{3/2}M_{Pl}.
\mu' \sim {\frac{\lambda' \lambda_{N}}{(16 \pi^{2})^{2}}} \Lambda_{Gravity}^2,
\end{equation}
where $\Lambda_{Gravity}\sim {\cal O}(10^{10}$ GeV) 
is the SUSY-breaking insertion. For $\lambda' \agt 10^{-8}$, 
this gives $(\mu' x)^{1/3} \agt 6 \  \tev$ thus
destabilizing the electroweak scale. In other words, to obtain  
\mbox{$\mu' x\sim {\cal O}(v^3)$}, one needs  $\lambda' \alt 10^{-11}$, 
in obvious contradiction with cosmology.

In our model, the operator that
can destabilize the hierarchy most severely is 
$\lambda' (H_{u} H_{d}) S^{2}/M_{Pl}$. The only field with a
messenger scale VEV is $S$. The lower bound on $\lambda'$ is now altered to
\begin{equation}
\lambda' \lambda_S \agt \left(\frac{T_* v_0}{v^2}\right)^2 \sim 10^{-3}
\end{equation}
for $v_0=y\sim 10^{5}$. 
With $\lambda_S\sim {\cal O}({10}^{-3})$, this leads
$\lambda'$ to the natural value $\lambda'\sim {\cal O}(1)$.
Such a large value of $\lambda'$ looks a lot worse than in 
the gravity-mediated case 
in terms of the destabilization of the hierarchy. 
However, a crucial observation is that in GMSB models 
the SUSY-breaking scale is typically very low.
In the model under consideration,
SUSY-breaking is communicated via a direct interaction 
$\xi_S S\overline\Phi\Phi$ and consequently 
the fundamental scale of SUSY-breaking
$F_0=F_S/\xi_S \sim \Lambda^2$ \cite{GMSB}, 
where $\Lambda$ is 50--100 TeV. As earlier, the operators
$\lambda' (H_{u} H_{d}) S^{2}/M_{Pl}$ and $\lambda_S H_{u} H_{d} S$
generate a term in the superpotential of the form $\mu' S$. 
In this case, 
\begin{equation}
\mu'  \sim {\frac{ \lambda' \lambda_{S}}{(16 \pi^{2})^{2}}} \Lambda^2
 \simeq \left({\frac{ \lambda' \lambda_{S}}{4\times 10^{-3}}}\right)
\left({\frac{\Lambda}{100\ {\rm TeV}}}\right)^2 {(30\ \rm GeV)}^2.
\end{equation}
Therefore, the stability of the electroweak scale can be maintained, 
$\mu' y\sim {\cal O}(v^3)$, while simultaneously satisfying the cosmological
constraints.
However, if the fundamental SUSY-breaking scale happens to be 
much higher $F_0\gg \Lambda^2$, the problem of destabilizing
the electroweak scale will remain.

\section{Discussion and Conclusion}

By design in our model, both $\mu$ and $B_\mu$ parameters are 
generated spontaneously at the right order, thus solving the
$\mu$-problem. The fine-tune for the cancellation
between $\mu^2$ and $m_{H_u}^2$ to obtain the correct
$M_Z$ is still at the percentage level, as in the MSSM.
The primary reason that we are able to overcome the $\mu$-problem
in GMSB \cite{mugmsb} is the inclusion of messenger-matter interactions 
in the superpotential. However, the smallness of the
parameters $\eta_S$ and $\lambda_S$ should be further
explored in terms of discrete symmetries arising from the SUSY-breaking 
sector.  

As for low-energy phenomenology, we see from Table~\ref{spectrum} 
that the sparticle spectrum is typical of gauge-mediated models. 
It is worth mentioning that if we require $k$ to be perturbative only below 
the messenger scale $\Lambda$, we find $k \leq 1.36$. This results in
the upper bound on the lightest Higgs boson mass to be about 170 GeV.

In summary, we studied in some detail the next-to-minimal supersymmetric 
standard model with the gauge mediation of supersymmetry breaking. 
We found that it is feasible to spontaneously generate the
$\mu$ and $B_\mu$ terms which are consistent with electroweak 
symmetry breaking and the current sparticle mass limits. 
Messenger sneutrinos with mass
in the range 6 to 25 TeV can serve as cold dark matter.
 It is also possible to find a solution to the cosmological domain wall
 problem consistent with EWSB due to the low scale of SUSY-breaking

\vskip 1.5cm
{\it Acknowledgments}: 
We thank H.-C. Cheng and M. Drees for helpful comments on the manuscript.
This work was supported in part by a DOE grant No. DE-FG02-95ER40896 
and in part by the Wisconsin Alumni Research Foundation.

\vskip 1.5cm
\appendix
\section{}
The 1-loop renormalization group equations of the couplings and soft-SUSY 
breaking terms relevant for our analysis are listed. We only
 include contributions from the third generation to the superpotential
and soft-terms.

\begin{eqnarray} 
W &=& \xi_S S \Phi \bar\Phi + \xi_N N \Phi \bar\Phi - {\eta_N\over2}
N^2 S - {\eta_S\over2} S^2 N + \lambda_N N H_u H_d  \nonumber\\
&&\quad {}+ \lambda_S S H_u H_d - {k\over3} N^3 + h_t H_u Q \bar U + h_b H_d Q 
\bar D + h_\tau H_d L \bar E,\\
V_{\rm soft} &=& m_{H_d}^2 |H_d|^2 + m_{H_u}^2 |H_u|^2 + m_N^2 |N|^2 +
m_{\tilde t_L}^2 |\tilde t_L|^2  +m_{\tilde t_R}^2 |\tilde t_R|^2+ m_{\tilde 
b_R}^2 |\tilde b_R|^2 +m_{\tilde \tau_L}^2 |\tilde \tau_L|^2
 \nonumber\\
&&\quad {}+m_{\tilde \tau_R}^2 |\tilde \tau_R|^2 + {1\over2} M_1 \tilde B \tilde B +  {1\over2} M_2 \tilde W 
\tilde W + {1\over2} M_3 \tilde g \tilde g + (A_t h_t \tilde t_L 
\tilde t_R H_u + A_b h_b \tilde b_L \tilde b_R H_d
\nonumber\\
&&\quad {} + A_{\tau} 
h_{\tau} \tilde \tau_L \tilde \tau_R H_d - \lambda_N A_{\lambda_N} H_d H_u N - 
{k\over3} A_k N^3 + h.c.) .
\end{eqnarray}
For the gauge and Yukawa couplings:
\begin{eqnarray}
16\pi^2 {d\over dt} g'  &=& 11g'^3,\quad 
16\pi^2 {d\over dt} g_2 = g_2^3,\quad
16\pi^2 {d\over dt} g_3 = - 3g_3^3,\\
16\pi^2 {d\over dt} h_t &=& h_t \left( 6h_t^2 + h_b^2 + \lambda_N^2 +
\lambda_S^2 - {13\over9}g'^2 - 3g_2^2 - {16\over3}g_3^2 \right), \\
16\pi^2 {d\over dt} h_b &=& h_b \left( 6h_b^2 + h_t^2 + h_\tau^2  + \lambda_N^2
+ \lambda_S^2 - {7\over9}g'^2 - 3g_2^2 - {16\over3}g_3^2 \right), \\
16\pi^2 {d\over dt} h_\tau &=& h_\tau \left( 4h_\tau^2 + 3h_b^2 + \lambda_S^2 -
3g'^2 - 3g_2^2 \right).
\end{eqnarray}
For the couplings particular to the NMSSM and to our model:
\begin{eqnarray} 
16\pi^2 {d\over dt} k\, \  &=& 6k ( k^2 + \lambda_N^2) - \eta_N^2 \left( 2k
+ \eta_S \right), \\
16\pi^2 {d\over dt} \lambda _N &=& \lambda_N \left( 2\lambda_S^2 + \eta_S^2 +
\eta_N^2 + 5\xi_N^2 + 4\lambda_N^2 + 2k^2 + 3h_t^2 + 3h_b^2 + h_\tau^2 - g'^2 -
3g_2^2 \right), \\
16\pi^2 {d\over dt} \lambda_S &=& \lambda_S \left( 2\lambda_N^2 + 4\lambda_S^2
+ \eta_S^2 + \eta_N^2 + 5\xi_S^2 + 3h_t^2 + 3h_b^2 + h_\tau^2 - g'^2 - 3g_2^2 
\right),\\
16\pi^2 {d\over dt} \xi_N &=& \xi_N \left( 9\xi_N^2 + 4\xi_S^2 + \eta_S^2 + 
\eta_N^2
+ 2\lambda_N^2 + 2k^2 + {13\over9}g'^2 - 3g_2^2 - {16\over3}g_3^2\right),\\
16\pi^2 {d\over dt} \xi_S &=& \xi_S \left( 9\xi_S^2 + 4\xi_N^2 + \eta_S^2 + 
\eta_N^2
+ 2\lambda_S^2 + {13\over9}g'^2 - 3g_2^2 - {16\over3}g_3^2\right),\\
16\pi^2 {d\over dt} \eta_N &=& \eta_N \left( 5\xi_S^2 + 10\xi_N^2 + 
3\eta_S^2 + 3\eta_N^2 +
2\lambda_S^2 + 4\lambda_N^2 + 4k^2 \right), \\
16\pi^2 {d\over dt} \eta_S &=& \eta_S \left( 5\xi_N^2 + 10\xi_S^2 + 
3\eta_S^2 + 3\eta_N^2 +
4\lambda_S^2 + 2\lambda_N^2 + 2k^2 \right).
\end{eqnarray}
For the trilinear couplings of the third generation:
\begin{eqnarray}
16\pi^2 {d\over dt} A_{t} &=& 12h_t^2 A_t +
2h_b^2 A_b + 2\lambda_N^2 A_\lambda - 4 \left( {13\over18}g'^2 M_1 + 
{3\over2}g_2^2 M_2 + {8\over3}
g_3^2 M_3 \right), \\
16\pi^2 {d\over dt} A_{b}  &=& 12h_b^2 A_b + 2h_t^2
 A_t + 2h_\tau^2 A_\tau + 2\lambda_N^2 A_\lambda- 4 
\left( {7\over18} g'^2 M_1 + 
{3\over2} g_2^2 M_2 + {8\over3}g_3^2 M_3 \right), \\
16\pi^2 {d\over dt}A_{\tau} &=& 8h_\tau^2 A_\tau + 6h_b^2
A_b + 2\lambda_N^2 A_\lambda - 6 \left( g'^2 M_1 + g_2^2 M_2 \right), \\
16\pi^2 {d\over dt} A_\lambda &=& 8\lambda_N^2 A_\lambda + 2\lambda_S^2
A_\lambda - 4k^2 A_k + 6h_t^2 A_t + 6h_b^2 A_b + 2h_\tau^2 - 2 
\left( g'^2 M_1 + 3g_2^2 M_2 \right), \\
16\pi^2 {d\over dt} A_k &=& 12 \left( k^2 A_k - \lambda_N^2 A_\lambda 
\right) + \eta_N^2 A_k.
\end{eqnarray} 
For the soft masses:
\begin{eqnarray}
16\pi^2 {d\over dt} M_1  &=& 22 g'^2 M_1 ,\quad 
16\pi^2 {d\over dt} M_2 = 2 g_2^2 M_2,\quad
16\pi^2 {d\over dt} M_3 = - 6 g_3^2 M_3,\\
16\pi^2 {d\over dt} m_{ \tilde t_L}^2 &=& 2 h_t^2 \left( m_{\tilde
t_L}^2 + m_{H_u}^2 + m_{\tilde t_R}^2 + A_t^2 \right) + 2 h_b^2 \left(
m_{\tilde t_L}^2 + m_{H_d}^2 + m_{\tilde b_R}^2 + A_b^2 \right)
\nonumber\\
&& \quad {} + {1\over3} g'^2\zeta-8 \left( {1\over36} g'^2 M_1^2 + {3\over4}
g_2^2 M_2^2 + {4\over3} g_3^2 M_3^2 \right), \\
16\pi^2 {d\over dt} m_{ \tilde t_R}^2 &=& 4 h_t^2 \left(
m_{\tilde t_L}^2 + m_{H_u}^2 + m_{\tilde t_R}^2 + A_t^2 \right) - 8 \left(
{4\over9} g'^2 M_1^2 + {4\over3} g_3^2  M_3^2 \right) - {4\over3} g'^2\zeta, \\
16\pi^2 {d\over dt} m_{ \tilde b_R}^2 &=& 4 h_b^2 \left(
m_{\tilde Q_3}^2 + m_{H_d}^2 + m_{\tilde b}^2 + A_b^2 \right) - 8 \left(
{1\over9} g'^2 M_1^2 + {4\over3} g_3^2 M_3^2 \right) + {2\over3} g'^2\zeta, \\
16\pi^2 {d\over dt} m_{ \tilde \tau_L}^2 &=& 2  h_\tau^2 \left( m_{\tilde
\tau_L}^2 + m_{H_d}^2 + m_{\tilde\tau_R}^2 + A_\tau^2 \right) - 8 \left( 
{1\over4}
g'^2 M_1^2 + {3\over4} g_2^2 M_2^2 \right) - g'^2\zeta, \\
16\pi^2 {d\over dt} m_{ \tilde \tau_R}^2 &=& 4  h_\tau^2 \left( m_{\tilde
\tau_L}^2 + m_{H_d}^2 + m_{\tilde\tau_R}^2 + A_\tau^2 \right) - 8 g'^2 M_1^2 + 
2g'^2\zeta, \\
16\pi^2 {d\over dt} m_{H_d}^2 &=& 6h_b^2 \left( m_{H_d}^2 + m_{\tilde t_L}^2 +
m_{\tilde b_R}^2 + A_b^2 \right) + 2\lambda_S^2 \left( m_{H_d}^2 + m_{H_u}^2
\right)\nonumber\\
&&\quad {} + 2h_\tau^2 \left( m_{H_d}^2 + m_{\tilde \tau_L}^2 + 
m_{\tilde\tau_R}^2 +
A_\tau^2 \right) + 2\lambda_N^2 \left( m_{H_d}^2 + m_{H_u}^2 + m_N^2 +
A_\lambda^2 \right) \nonumber\\
&&\quad {}- 8 \left( {1\over4} g'^2 M_1^2 + {3\over 4} g_2^2 M_2^2 \right) -
g'^2\zeta, \\
16\pi^2 {d\over dt} m_{H_u}^2 &=& 6h_t^2 \left( m_{H_u}^2 + m_{\tilde t_L}^2 +
m_{\tilde t_R}^2 + A_t^2 \right) + 2\lambda_S^2 \left( m_{H_d}^2 + m_{H_u}^2
\right) \nonumber\\
&&\quad - 8 \left( {1\over4} g'^2 M_1^2 + {3\over4} g_2^2 M_2^2 \right)
 + 2\lambda_N^2 \left( m_{H_d}^2 + m_{H_u}^2 + m_N^2 + A_\lambda^2
\right) + g'^2\zeta, \\
16\pi^2 {d\over dt} m_N^2 &=& 4\lambda_N^2 \left( m_N^2 + m_{H_u}^2 + m_{H_d}^2
+ A_\lambda^2 \right) + 4k^2 \left( 3m_N^2 + A_k^2 \right),
\end{eqnarray}
where $\zeta=\sum_{i} Y_{i}m_{i}^{2}$ is the hypercharge-weighted sum of the 
squares of the soft masses. As a consequence of the 
 boundary conditions at the messenger scale,  $\zeta=0$ at all 
energies. 

\section{}

The squared mass matrices for the neutral and 
charged Higgses are 
presented. The analytic expressions for their diagonalization are displayed.
\begin{equation}
M^{2}_{CP-even}= {\frac{1}{2}}{\frac{\partial^{2} V^{1-loop}_{neutral}}
{\partial 
v_{i}\partial v_{j}}} ={\frac{1}{2}} \left( \begin{array}{ccc}
 a & d & e \\
 d & b & f \\
 e & f & c
\end{array} \right),
\end{equation}
where $v_{1}=v_{d},\  v_{2}=v_{u},\   v_{3}=x$ and
\begin{eqnarray}
a  &=&  {\frac{1}{ v_{d}}} ( \bar{g}^{2} v_{d}^{3} + 
  {v_u} ( ( 2 k {x^2} + 
         2xA_{\lambda_{N}}+
        2 x y {{\eta }_N} + {y^2} {{\eta }_S} )  
      {{\lambda }_N} + ( -2 {F_S} + {x^2} {{\eta }_N} + 
        2 x y {{\eta }_S} ) {{\lambda }_S} ) ) \nonumber\\ 
&&-(\partial_{v_d} \Delta V)/{v_d}+\partial_{{v_d},\,{v_d}}^{2} \Delta V \\
b  &=&  {\frac{1}{ v_{u}}} ( \bar{g}^{2} v_{u}^{3} + 
  {v_d} ( ( 2 k {x^2} + 
         2xA_{\lambda_{N}}+
        2 x y {{\eta }_N} + {y^2} {{\eta }_S} )  
      {{\lambda }_N} + ( -2 {F_S} + {x^2} {{\eta }_N} + 
        2 x y {{\eta }_S} ) {{\lambda }_S} ) ) \nonumber\\
&&-(\partial_{v_u} \Delta V)/{v_u}+\partial_{{v_u},\,{v_u}}^{2} \Delta V \\
c  &=&  {\frac{1}{8\pi^{2} x y}} (-16 k {{\pi }^2} {x^2} y {A_k} - 
  F_{S}^{2} {{\xi }_N} {{\xi }_S} + 
  16 {{\pi }^2} {y^2} {F_S} {{\eta }_S} + 
  8 {{\pi }^2} y (  
     y {{\eta }_N}  ( 6 k {x^2}  + 
        ( 3 {x^2} - {y^2} )  {{\eta }_S} 
      \nonumber\\
&& +  2 {v_d} {v_u} {{\lambda }_N} )  + 2 {x^3} \eta_{N}^{2} +
     2 ( 4 {k^2} {x^3} + 
     A_{\lambda_{N}}   {v_d} {v_u} 
         {{\lambda }_N} + 
        y {v_d} {v_u} {{\eta }_S} {{\lambda }_S} - 
        y v_{d}^{2} {{\lambda }_N} {{\lambda }_S} - 
        y  v_{u}^{2} {{\lambda }_N} {{\lambda }_S} 
      ))) \nonumber\\ 
&&-(\partial_{x} \Delta V)/x+\partial_{{x},\,{x}}^{2} \Delta V \\
d  &= &  - 
  ( 2 k {x^2} +  2xA_{\lambda_{N}}+
     2 x y {{\eta }_N} + {y^2} {{\eta }_S} )  
   {{\lambda }_N}  + 
  {{\lambda }_S} ( 2 {F_S}- {x^2} {{\eta }_N} - 
     2 x y {{\eta }_S} + 4 {v_d} {v_u} {{\lambda }_S}
      )\nonumber\\
 && - \bar{g}^{2} {v_d} {v_u} +  4 {v_d} {v_u} \lambda_{N}^{2} 
+\partial_{{v_d},\,{v_u}}^{2} \Delta V \\
e  &=&  4 {v_d} {{\lambda }_N} 
   ( x {{\lambda }_N} + y {{\lambda }_S} )  - 
 2 {v_u} ( ( 2 k x +  A_{\lambda_{N}} + 
        y {{\eta }_N} )  {{\lambda }_N} + 
     ( x {{\eta }_N} + y {{\eta }_S} )  
      {{\lambda }_S} )+\partial_{{v_d},\,{x}}^{2} \Delta V  \\
f  &=&  4 {v_u} {{\lambda }_N} 
   ( x {{\lambda }_N} + y {{\lambda }_S} )  - 
 2 {v_d} ( ( 2 k x + A_{\lambda_{N}}  + 
        y {{\eta }_N} )  {{\lambda }_N} + 
     ( x {{\eta }_N} + y {{\eta }_S} )  
      {{\lambda }_S} )+\partial_{{v_u},\,{x}}^{2} \Delta V .
\end{eqnarray}
The eigenvalue equation from this matrix is
\begin{equation}
{(m^{2})}^{3}+a_{2}{(m^{2})}^{2}+a_{1}m^{2}+a_{0} = 0
\end{equation}
where 
\begin{eqnarray}
a_{0} & = &- a b c  - 
  2 de f   + a {f^2}+ b {e^2}+ c {d^2}  \\
a_{1} & = &a b + a c + b c - ({d^2} + {e^2} + {f^2}) \\ 
a_{2} & = & -(a+b+c).
\end{eqnarray}
With
\begin{equation}
r =  {\frac{1}{6}}(a_{1}a_{2}-3 a_{0})-{\frac{1}{27}}a_{2}^{3}, \quad
q =  -{\frac{1}{3}}a_{1}+{\frac{1}{9}}a_{2}^{2}, \quad
\theta  =  {\frac{1}{3}}\arccos {\frac{r}{q^{3/2}}} ,
\end{equation}
the eigenvalues are
\begin{eqnarray}
m_{S_{1}}^{2} & = & -\sqrt{q}(\cos\theta+\sqrt{3}\sin\theta)-{\frac{a_{2}}{3}} 
\\
m_{S_{2}}^{2} & = & \sqrt{q}(-\cos\theta+\sqrt{3}\sin\theta)-{\frac{a_{2}}{3}} 
\\
m_{S_{3}}^{2} & = & 2\sqrt{q}\cos\theta- {\frac{a_{2}}{3}}.
\end{eqnarray}
Now for the CP-odd Higgs bosons. The mass squared matrix is
\begin{equation}
M^{2}_{CP-odd}=  \left( \begin{array}{cc}
 r & j  \\
 j & s  \\
\end{array} \right)
\end{equation}
where
\[ 
\begin{array}{l}
r={\frac{ {v^2}}{2 {v_d} {v_u}}} \left( \left( 2 k {x^2} + 2xA_{\lambda_{N}}+
          2 x y {{\eta }_N} + {y^2} {{\eta }_S} \right)  
        {{\lambda }_N} + \left( -2 {F_S} + 
          x \left( x {{\eta }_N} + 2 y {{\eta }_S} \right) 
           \right)  {{\lambda }_S} \right) \\
\\
s={\frac{1}{16 {{\pi }^2} x y}}
 [48 k {{\pi }^2} {x^2} y {A_k} -
  F_{S}^{2} {{\xi }_N} {{\xi }_S} + 
  16 {{\pi }^2} y {F_S} 
   \left( 2 x {{\eta }_N} + y  {{\eta }_S} \right)-8 {{\pi }^2} y
 \Bigl( {{\eta }_N} 
     \Bigl( 2 k {x^2} y + 
       y \left( {x^2} + {y^2} \right)  {{\eta }_S} \\
 -2 {v_d} {v_u} \left( y {{\lambda }_N} + 
          2 x {{\lambda }_S} \right)  \Bigr)  + 
    2 \left( y {{\eta }_S} 
        \left( 2 k x y - {v_d} {v_u} {{\lambda }_S} \right)
           + {{\lambda }_N} 
        \left( \left( -4 k x -A_{\lambda_{N}} 
              \right)  {v_d} {v_u} + 
          y {{v}^2} {{\lambda }_S}  \right)  \right) 
     \Bigr) ] \\
 \\
j=-\left( v \left( \left( 2 k x - A_{\lambda_{N}} + 
         y {{\eta }_N} \right)  {{\lambda }_N} + 
      \left( x {{\eta }_N} + y {{\eta }_S} \right)  
       {{\lambda }_S} \right)  \right). \\ 
\end{array}
\]
The transformation matrix that takes us from the $(A^{0} \equiv \sin\beta\, 
{\rm Im}H_{d}+\cos\beta\,{\rm Im}H_{u})-{\rm Im}N$ basis to the diagonal basis is 
\begin{equation}
\left( \begin{array}{cc} \cos\gamma & \sin\gamma \\ -\sin\gamma & \cos\gamma 
\end{array} \right)
\end{equation}
where
\begin{equation}
\sin 2\gamma={\frac{-2 j}{{\sqrt{{{\left(  s-r \right) }^2} + 4 {j^2}}}}}, 
\qquad
\cos 2\gamma={\frac{s-r}{{\sqrt{{{\left(  s-r \right) }^2} + 4  {j^2}}}}},
\end{equation}
and the eigenvalues are
\begin{equation}
m_{P_{1},P_{2}}^{2}={\frac{1}{2}}(s+r \mp \sqrt{(s-r)^{2}+4j^{2}}).
\end{equation}
Finally, the squared mass of the charged Higgs is
\begin{equation}
m_{H^\pm}^{2} =  M_W^2- (\lambda_N^2+\lambda_S^2)v^2 + r
\end{equation}
where $r$ is the first entry of the CP-odd Higgs matrix.

\end{document}